\documentclass{appolb}
\usepackage{graphicx}
\usepackage{mathptmx}      
\usepackage{flushend}
\usepackage[numbers,sort&compress]{natbib}
\usepackage[colorlinks,citecolor=blue,urlcolor=blue,linkcolor=blue]{hyperref}
\usepackage{feynmp}
\usepackage{cancel}
\usepackage{adjustbox}
\usepackage{array}
\usepackage{multirow}
\usepackage{bm}
\usepackage[scaled=1.15]{urwchancal}
\usepackage{tikz}
\usetikzlibrary{patterns}
\usetikzlibrary{plotmarks}
\begin{document}
\title{Resonance Production of  Excited u-quark at FCC Based $\gamma p$ Colliders%
}
\author{Yusuf Oguzhan G\"unaydin
\address{Department of Physics, Kahramanmaras Sutcu Imam University, Kahramanmaras, Turkey}
\\
{
}
Mehmet Sahin
\address{Department of Physics, Usak University, Usak, Turkey}
\\
{
}
Saleh Sultansoy 
\address{TOBB University of Economics and Technology, Ankara, Turkey \\ ANAS Institute of Physics, Baku, Azerbaijan}
}

\maketitle
\begin{abstract}
Several Beyond the Standard Model theories are proposed that fermions might have composite substructure. The existence of excited quarks is going to  be the noticeable proof for  the compositeness of Standard Model fermions.  For this reason, excited quarks have been investigated  by  phenomenological and experimental high energy physicists  at  various collider options for the last few decades. The Future Circular Collider (FCC) has been recently planned as particle accelerator to be established at CERN. Beside the $\sqrt{s}$ = 100 TeV proton-proton collisions, the FCC includes electron-positron and electron-proton collision options. Construction of linear $e^- e^+$ colliders (or dedicated e-linac) tangential to the FCC will afford an opportunity to handle multi-TeV $ep$ and $\gamma p$ collisions. In this respect, we executed a simulation of the resonance production of the excited $u$ quark at the FCC based $\gamma p$ colliders with choosing both  the polarized and unpolarized photon beams. The findings revealed that the  chirality structure of the $q^*$-$q$-$\gamma$ vertex can be determined by the photon beam polarization. The attainable mass limits of the excited $u$ quark reached the highest values when the polarized photon beam was chosen. In addition, the ultimate compositeness scale values can be handled by appropriate choice of  the photon beam polarization. 
\end{abstract}
  
\section{\label{sec:int}Introduction}
The Standard Model (SM), the most reliable theory in particle physics,  shows incredible consistency with experiments and reaches its last prediction after the CMS and the ATLAS collaborations, which both declared the detection of the Higgs boson \cite{cms2012, atlas2012} in 2012.
Despite the marvelous success of the SM on a wide range of phenomena in particle physics, there remains unsolved mysteries that the SM does not explain. The quark-lepton symmetry, family replication, charge quantization,  plenty numbers of elementary particles, parameters and the like are unsolved issues  in the SM frame. Therefore, numerous models are proposed in an attempt to answer the afore-mentioned problems. One of  these approaches; namely, compositeness has an assumption that SM fermions are compound states of more fundamental particles; preons \cite{pati1974,dsouza1992}. Numerous preonic models have been suggested by particle physicists for more than forty years \cite{shupe1979, harari1979,terazawa1980,terazawa1982,terazawa1983,eichten1983, fritzsch1981, celikel1998, desouza2008, fritzsch2016}. Due to preonic interactions caused by preon models, plenty of  new types of particles are expected, such as  excited quarks and leptons, leptoquarks, leptogluons, diquarks, color sextet quarks, dileptons and so on.

Excited fermions are comprised of  excited quarks ($q^*$) and leptons ($l^*$) that can be considered as the excited state of  SM fermions. They could have spin-1/2 and spin-3/2 states and their masses are expected much heavier than  SM fermions. As a result, the discovery of excited fermions will be a direct proof of  SM quarks' and leptons' compositeness. After the first publication about excited leptons which was written in 1965 \cite{low1965}, scores of theoretical, phenomenological \cite{renard1983,kuhn1984, pancheri1984, rujula1984, hagiwara1985, kuhn1985, baur1987, spira1989, baur1990, boudjema1993,cakir1999,cakir2000,cakir2001, eboli2002, cakir2004, ccakir2004,cakir2008,caliskan2017, caliskan2017exc} and experimental \cite{cdf1995,h1_2000,l3_2000,chekanov2002,abdallah2006,cms2014,atlas2016photon,atlas2016,atlas2017,cms2016,cms2016photon,cms2017,cms2017eej} researchers has been focused on  proving the existence of  excited fermions. The historical development of fundamental blocks of the matter \cite{sahin2011} shows that new substructures of elementary particles are discovered by new experimental findings and this evidence attracts attention of  particle physicists to do research on  excited quarks and leptons.       

Excited quarks decay into four final states with light jets of ($q^*\rightarrow jj$), ($ q^*\rightarrow j\gamma$), ($q^*\rightarrow jW$), and ($q^*\rightarrow jZ$). The most recent experimental results regarding excited quark mass are provided by the CMS and ATLAS collaborations \cite{cms2016, cms2016photon,cms2017,cms2017eej,  atlas2016, atlas2016photon,atlas2017,pdg2016}. $q^*$ mass exclusion limits are $m_{q^*} = 6.0$ TeV for $q^*\rightarrow jj$, $m_{q^*} = 5.5$ TeV  for $ q^*\rightarrow j\gamma$, $m_{q^*} = 3.2 $ TeV  for $ q^*\rightarrow jW$ and $m_{q^*} = 2.9$ TeV  for $ q^*\rightarrow jZ$. For these experimental limits on excited quark mass, compositeness scale ($\Lambda$) is considered to be equal to $m_{q^*}$.

In this paper, we investigate resonant production of the  up-type excited quark ($u^*$) with dijet final state at two different center of mass (CM) energies \cite{acar2017} of the Future Circular Collider (FCC) \cite{fcc2014} based $\gamma p$-colliders \cite{ciftci1995}. In addition, we neglect possible contact interactions at this stage. We present the FCC based colliders  options and their parameters, specifically $\gamma p$-colliders in the section \ref{sec:II}, $q^*$ effective interaction Lagrangian and decay width  in the section \ref{sec:III} and leading order production cross sections and signal-background analysis using unpolarized and polarized photons in the section \ref{sec:IV}. Finally, outcomes of the  $u^*$ mass limitations, the  compositeness scale ($\Lambda$)  inquiries and  interpretation of our findings are presented in the last section.  

\section{\label{sec:II}FCC based $\gamma p$-colliders}

\begin{figure}[h!]
	\centering
	\includegraphics[scale=0.3]{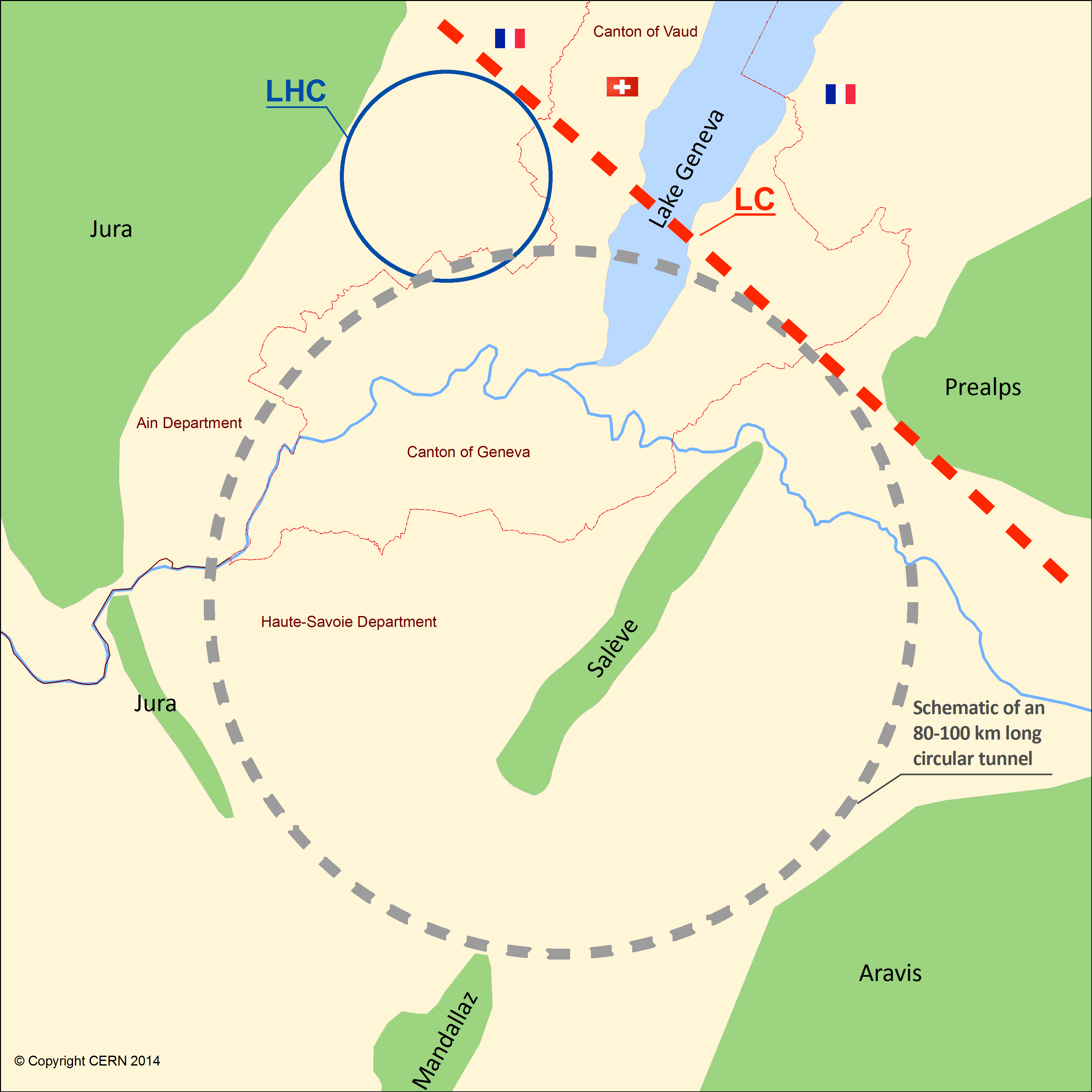}%
	\caption{\label{fig:fcc} Schematic drawing of the Future Circular Collider and the Linear Collider}
\end{figure}

Throughout the last 40 years of the particle  accelerator development, several groups and collaborations proposed linac-ring type colliders  (see reviews \cite{sultanov1989, wiik1993, brinkmann1997,sultansoy1998, sultansoy1999, sultansoy2004, akay2010}). Concerning energy frontier lepton-hadron options, VLEPP+UNK, THERA and  LHeC were proposed in the 1980s, 1990s and 2000s, respectively. The latter option \cite{abelleira2012}  is planned to be established at CERN around the 2020s. Furthermore, after the Large Hadron Collider (LHC) physics program are completed, the FCC \cite{benedikt2015future}  will be seen as  experimental particle physics frontier machine by  the high energy physics community.  The FCC is planned nearly 4 times bigger circumferences (Figure \ref{fig:fcc})  and about 7 times higher center of mass energy than the LHC. The FCC is considered as three options; (1) the electron-positron (FCC-ee) \cite{wenninger2014future}, (2) the proton-proton (FCC-pp) \cite{wenninger2014} and (3) the electron-proton (FCC-ep) \cite{benedikt2015future} colliders.  To measure new findings with high precision, FCC-ee is an appropriate  option, notwithstanding,  FCC-pp and  FCC-ep are needed for deep investigation of interactions. That is, many features of the Higgs boson can be measured by  FCC-ee whose collision energy varies  between  91 and 350 GeV.  However, further measurements like Higgs self-interactions and top quark Higgs bosons interaction could be achieved by FCC-pp at 100 TeV center of mass energy. Besides,  quark substructure discovery might be happen at the FCC-ep collider. 

With respect to excited quark, we focused on the FCC based $\gamma p$ collider within the scope of this study. There are several options for lepton-hadron  collision but we preferred the FCC based electron-proton colliders by using International Linear Collider (ILC) and Plasma Wakefield Accelerator-Linear Collider (PWFA-LC) \cite{acar2017}. In addition to the FCC based $ep$ colliders, $\gamma p$ colliders  \cite{ciftci1995, aksakal2007} could be utilized by exploiting Compton backscattering \cite{ginzburg1983,ginzburg1984,telnov1990}. Main parameters of the $ep$ and $\gamma p$  colliders which we used in our calculations are listed in Table \ref{tab:lcpar}.

\begin{table} [h!]
	\caption{\label{tab:lcpar} Energy and luminosity parameters of the ILC$\otimes$FCC and PWFA-LC$\otimes$FCC based $ep$ and $\gamma p$ colliders }
	\begin{tabular*}{\columnwidth}{@{\extracolsep{\fill}}llllll@{}} 	\hline	
		Collider Name&\begin{tabular}[c]{@{}l@{}}E$_e$ \\ $(TeV)$\end{tabular}&\begin{tabular}[c]{@{}l@{}}E$_{\gamma}^{max}$\\$(TeV)$\end{tabular}&\begin{tabular}[c]{@{}l@{}}$\sqrt{s}_{ep}$\\$(TeV)$\end{tabular}&\begin{tabular}[c]{@{}l@{}}$\sqrt{s}_{\gamma p}^{max}$\\$(TeV)$\end{tabular}&\begin{tabular}[c]{@{}l@{}}$\mathcal{L}_{int}$\\$(fb^{-1}/year)$\end{tabular} \\ \hline
		ILC$\otimes$FCC& 0.5 & 0.42 &10&9.1& 10-100 \\ \hline
		PWFA-LC$\otimes$FCC&  5 & 4.15& 31.6 & 28.8 & 1-10 \\ \hline
	\end{tabular*}
\end{table}

\section{\label{sec:III}Spin-1/2 excited quark interaction Lagrangian and decay width}

Interaction between spin-1/2 excited quarks, SM quarks and gauge bosons is described by  the magnetic type effective Lagrangian \cite{kuhn1984,rujula1984,baur1987, pdg2016} : 
\begin{equation}
\label{eq:intLag}
{L_{eff} = \frac{1}{2\Lambda}\bar{q^*}\sigma^{\mu\nu}[g_s f_s \frac{\lambda^a}{2}F^a_{\mu \nu} + g f \frac{\vec{\tau}}{2}\vec{W_{\mu \nu}} + g'f' \frac{Y}{2}B_{\mu \nu}] (\eta_{L}\frac{1-\gamma_5}{2} + \eta_{R}\frac{1+\gamma_5}{2})q + H.c. }.
\end{equation}
As illustrated above, $\Lambda$  denotes compositeness scale, $q^*$ and $q$ represent  spin-1/2 excited quark and ground state quark   respectively, $F^a_{\mu \nu}$, $\vec{W_{\mu \nu}}$, $B_{\mu \nu}$ are the field strength tensors for gluon, SU(2) and U(1), $\lambda^a$ are 3$\times$3 Gell-Mann  matrices,  $\vec{\tau}$ is the Pauli spin matrices, $Y =1/3$  is weak hypercharge, $g_s,\;g,\;g'$ are gauge coupling constants, $f_s$, $f$ and $f'$ are  free parameters that are chosen equal to 1.  $\eta_{L}$  and  $\eta_{R}$  are the left-handed and  the right-handed chirality factors, respectively. The couplings $\eta_{L/R}$ are uniquely defined by the gauge-group representation of the excited states: $\eta_L$ is only possible if the right-handed excited quarks are isospin doublets, while $\eta_R$ is only possible if the left-handed excited quarks are isospin singlets. The normalization of the coupling  was chosen such that  $max(|\eta_{L}|, |\eta_{R}|) =1 $ and chirality conservation requires $\eta_{L}\eta_{R} = 0$ \cite{pdg2016}.

\begin{figure} [h!]
	\centering
	\includegraphics[width=0.5\columnwidth]{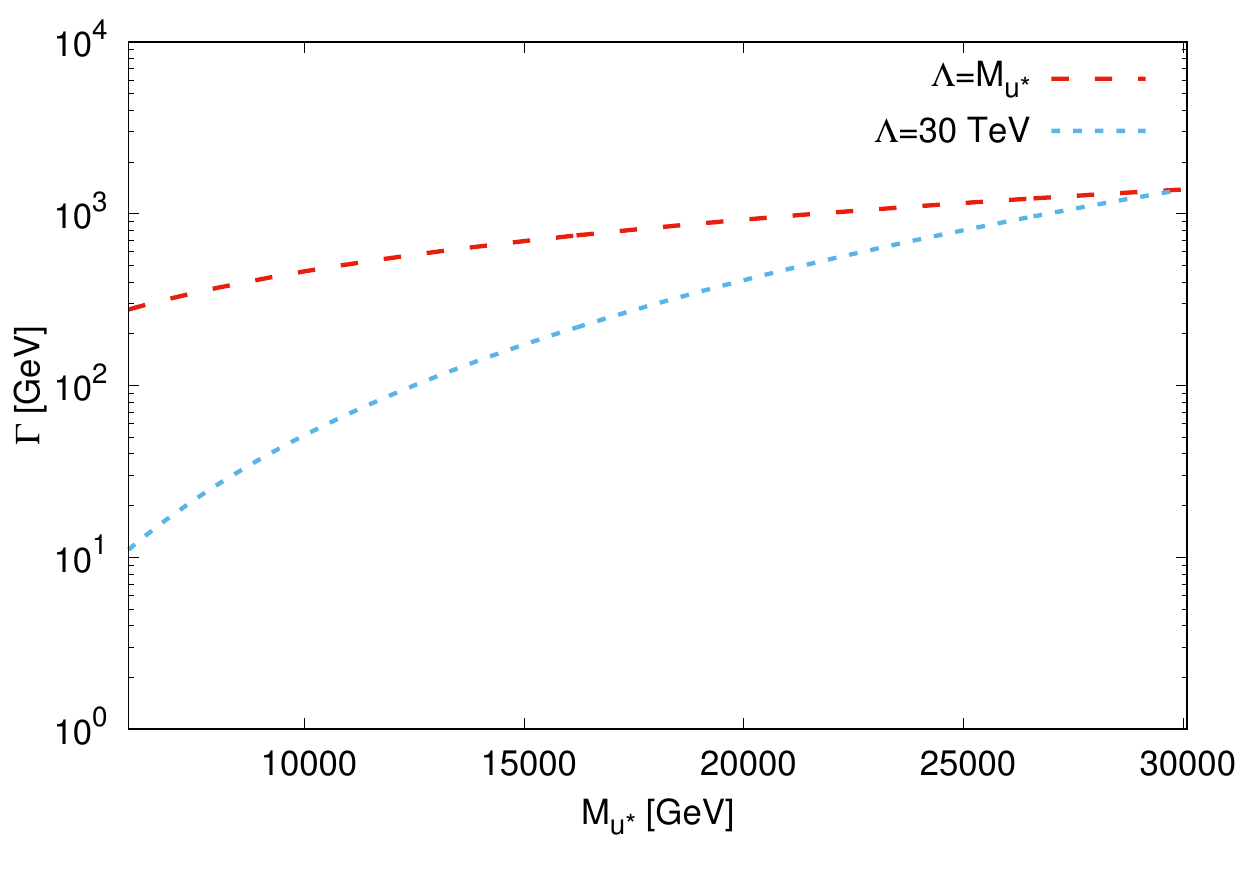}
	\caption{$u^*$ decay width correlations with excited quark mass  at $u^*$ mass equals  compositeness scale and $\Lambda = 30$ TeV}
	\label{fig:decay} 
\end{figure}

We implemented this interaction Lagrangian into CalcHEP software \cite{calchep2013} by using LanHEP \cite{LanHEP,lanhep2016}.	As we earlier mention in Section \ref{sec:int}, there are four decay channels for $q^*$  and we plotted total decay width with respect to the excited quark mass by taking compositeness scale equals  $q^*$ mass  and $\Lambda = 30$ TeV in  Figure \ref{fig:decay}.  It is illustrated that excited quark mass values are correlated with decay widths positively.

\section{\label{sec:IV}Excited $u$-quark production via proton collisions with unpolarized and polarized  photon at  $\sqrt{s}_{\gamma p}$ = 9.1 and 28.8 {TeV} }

\begin{figure}[h!]
	\centering
	\includegraphics[width=0.5\columnwidth]{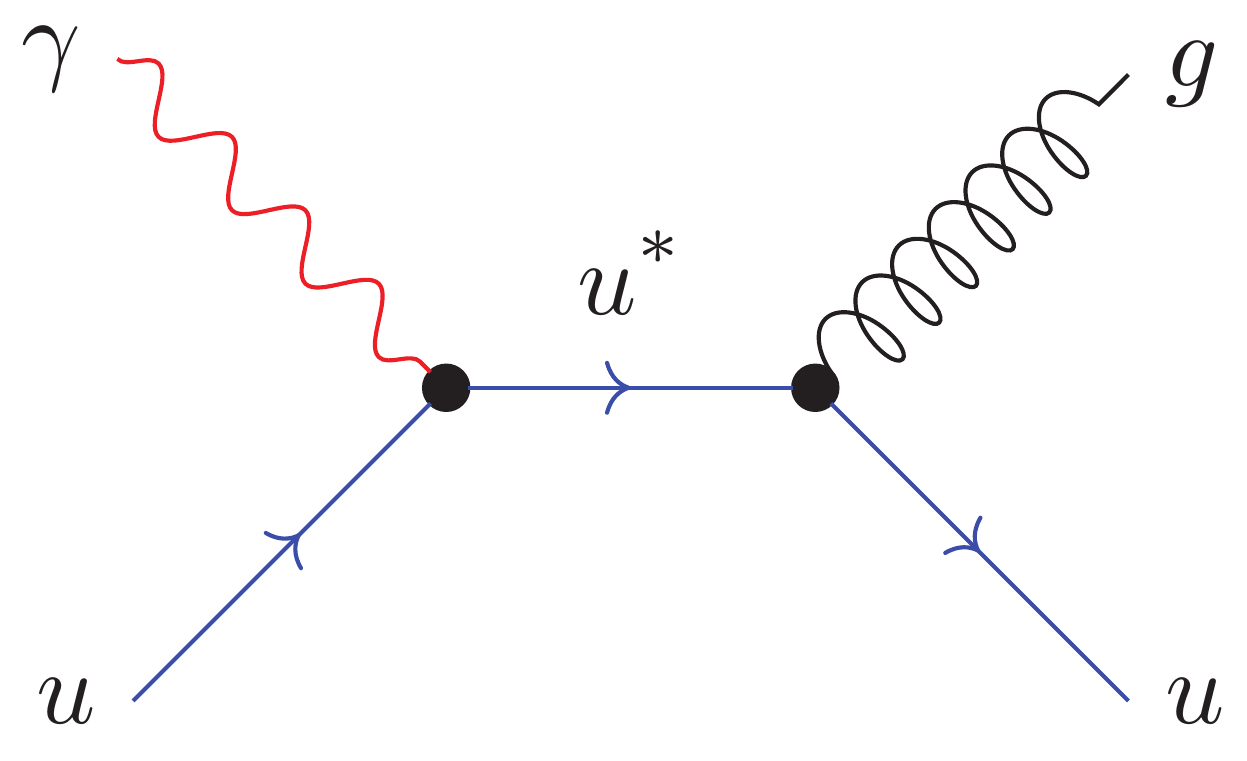}%
	\caption{Feynman diagram for signal.}
	\label{fig:feynsignal} 
\end{figure}

In our calculation, we used two types of particle beams; proton and photon (see Section \ref{sec:II}). 50 TeV proton beam comes from the FCC and we chose  CTEQ6L quark distribution function \cite{pumplin2002,stump2003}  with factorization and renormalization scales equal to $M_{u^*}$  in numerical calculations.  On the other hand, we had polarized and unpolarized high energy photon beams \cite{borden1993,dangelo2000}, which were obtained from Compton backscattering \cite{ginzburg1983,ginzburg1984,telnov1990} of laser beam  on ILC or PWFA-LC electrons.  The Feynman diagram for resonant production of $u^*$ in photon-proton collisions is presented in Figure \ref{fig:feynsignal}.   

\subsection{\label{subsec:cross}Cross Sections} 

In numerical calculations, we chose $\eta_{L} = 1$, $\eta_{R} = 0$ option for interaction Lagrangian   (Eq. \ref{eq:intLag}). Following, we inserted corresponding  electron and proton energies and chose laser photon option which corresponds to Compton backscattering photons  in CalcHEP framework. The energy spectrum of  backscattered laser photons, we used, is given in Refs. \cite{ginzburg1983,ginzburg1984,telnov1990,pukhov2012calchep} with a detailed explanation.

\begin{figure}[h!]
	\centering
	
	\includegraphics[width=0.49\columnwidth]{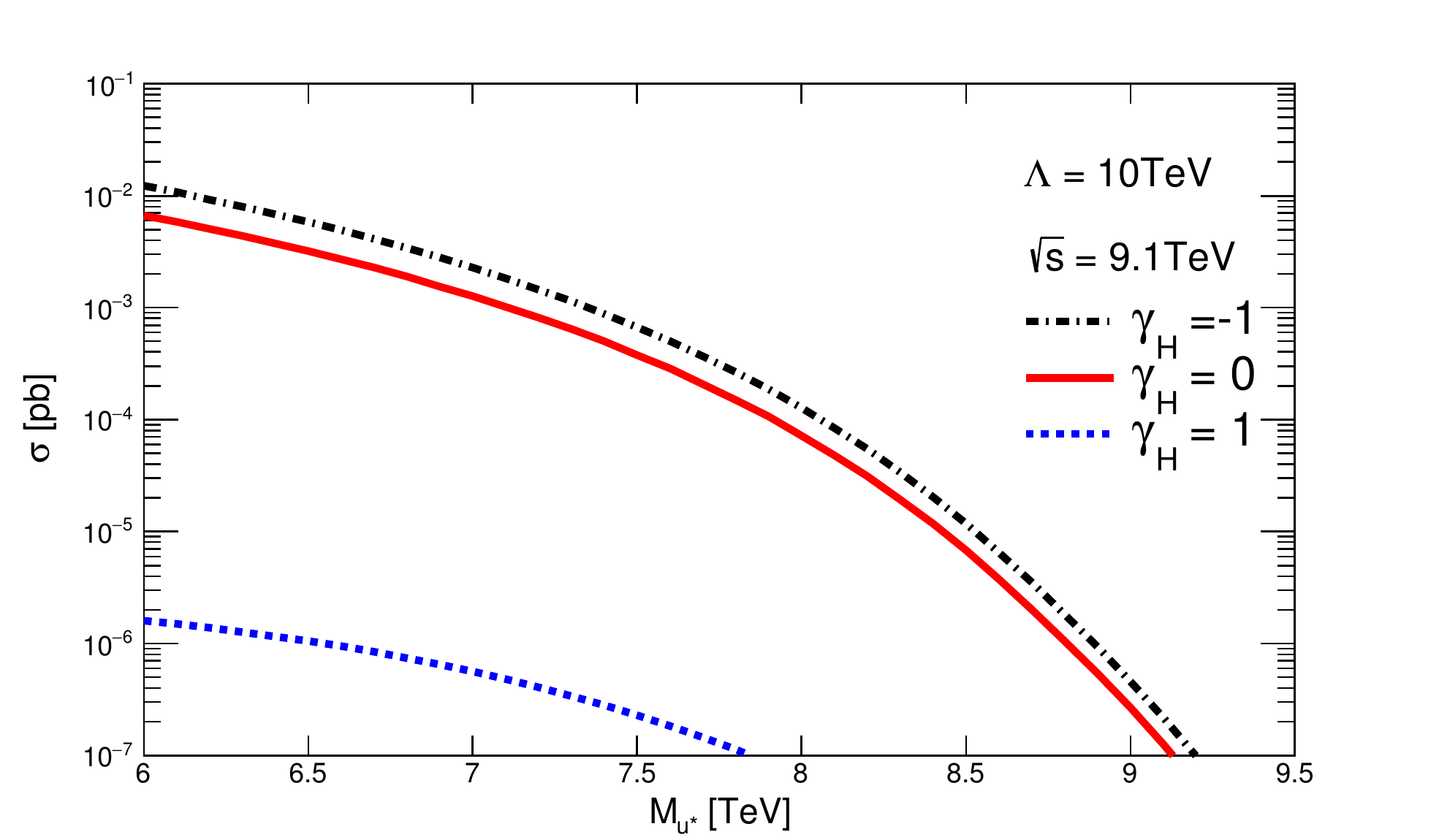}
	\includegraphics[width=0.49\columnwidth]{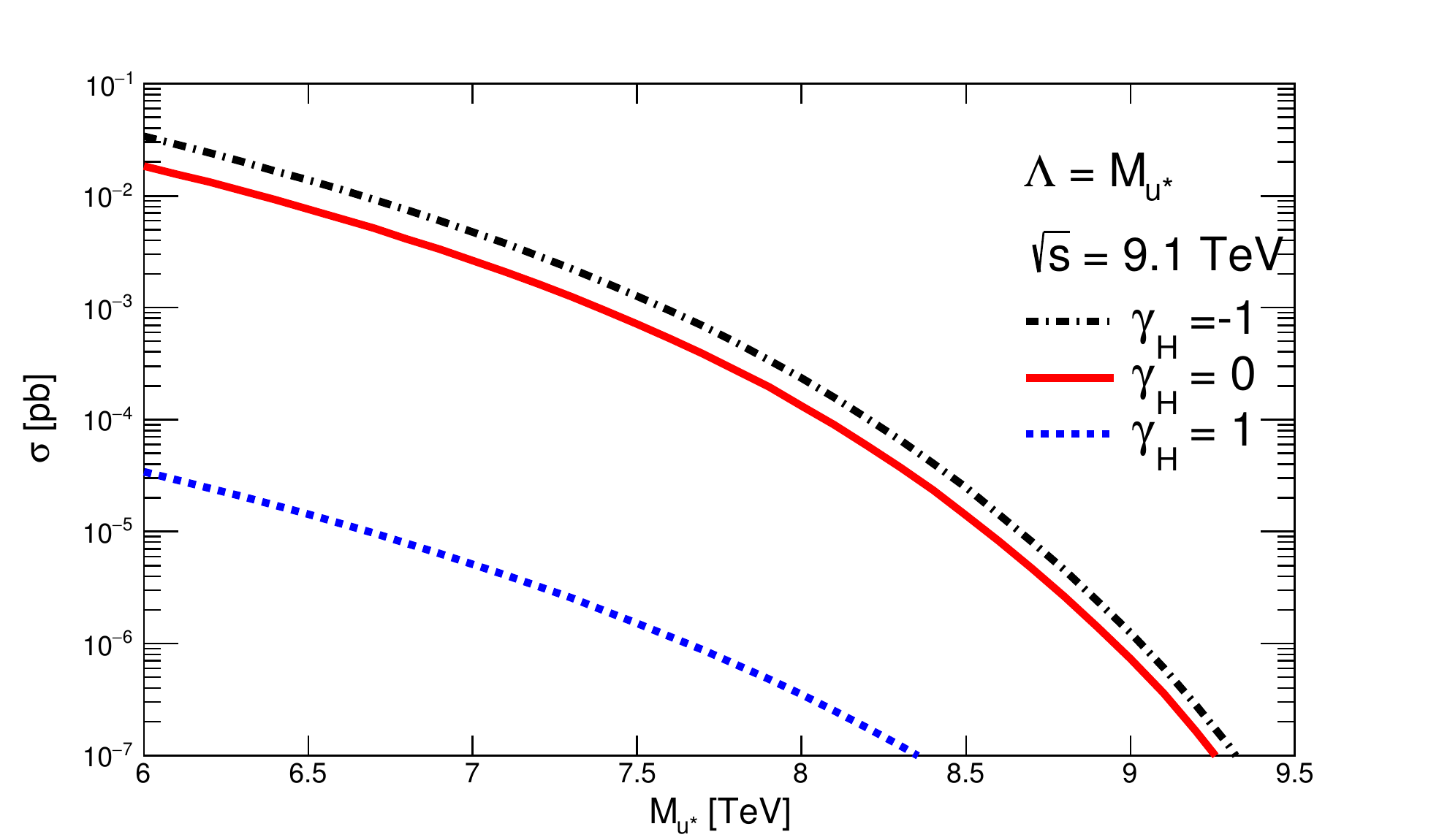}
	
	\caption{\label{fig:sc9100} $u^*$ cross section values  with respect to  its mass  for proton collision with both polarized and unpolarized photon beams at $\sqrt{s} = 9.1$ TeV.  On the left panel compositeness scale was chosen as 10 TeV and on the right panel $u^*$ mass was taken the same as compositeness scale.}
\end{figure}

Figure \ref{fig:sc9100} shows the cross section values with respect to $u^*$ mass for polarized (helicity $\gamma_{\mathcal{H}}$  equals 1 and -1) and unpolarized ($\gamma_{\mathcal{H}}$  = 0) photon beams colliding with proton beam at 9.1 TeV center of mass energy.  It is seen that excited quark could be produced with sufficiently high  cross section up to roughly 8 TeV both for $\Lambda = 10$ TeV and $\Lambda = M_{u^*}$, corresponding to 10 events for 100 $fb^{-1}$ luminosity value.

\begin{figure}[h!]
	\centering
	\includegraphics[width=0.5\columnwidth]{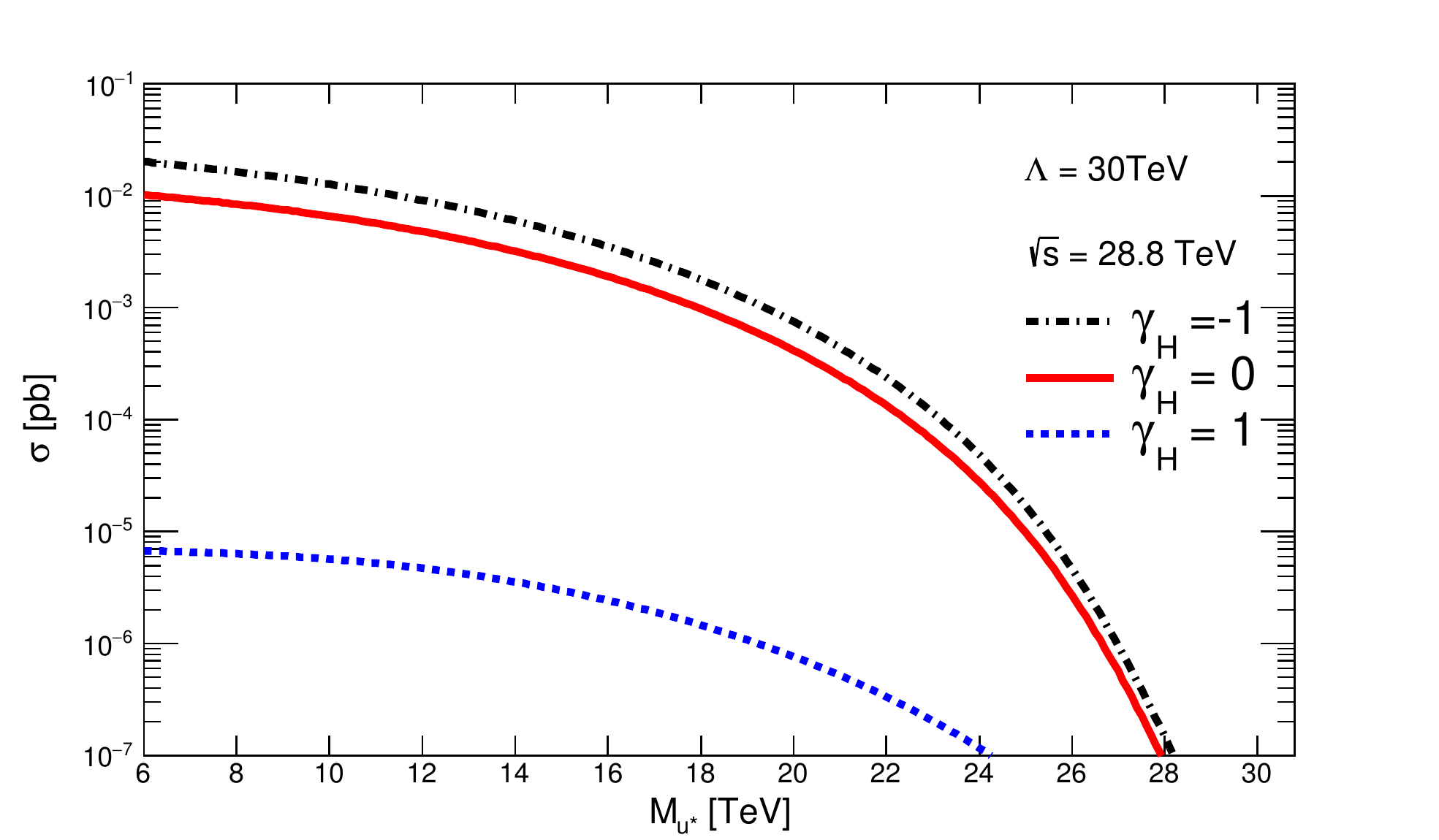}%
	\includegraphics[width=0.5\columnwidth]{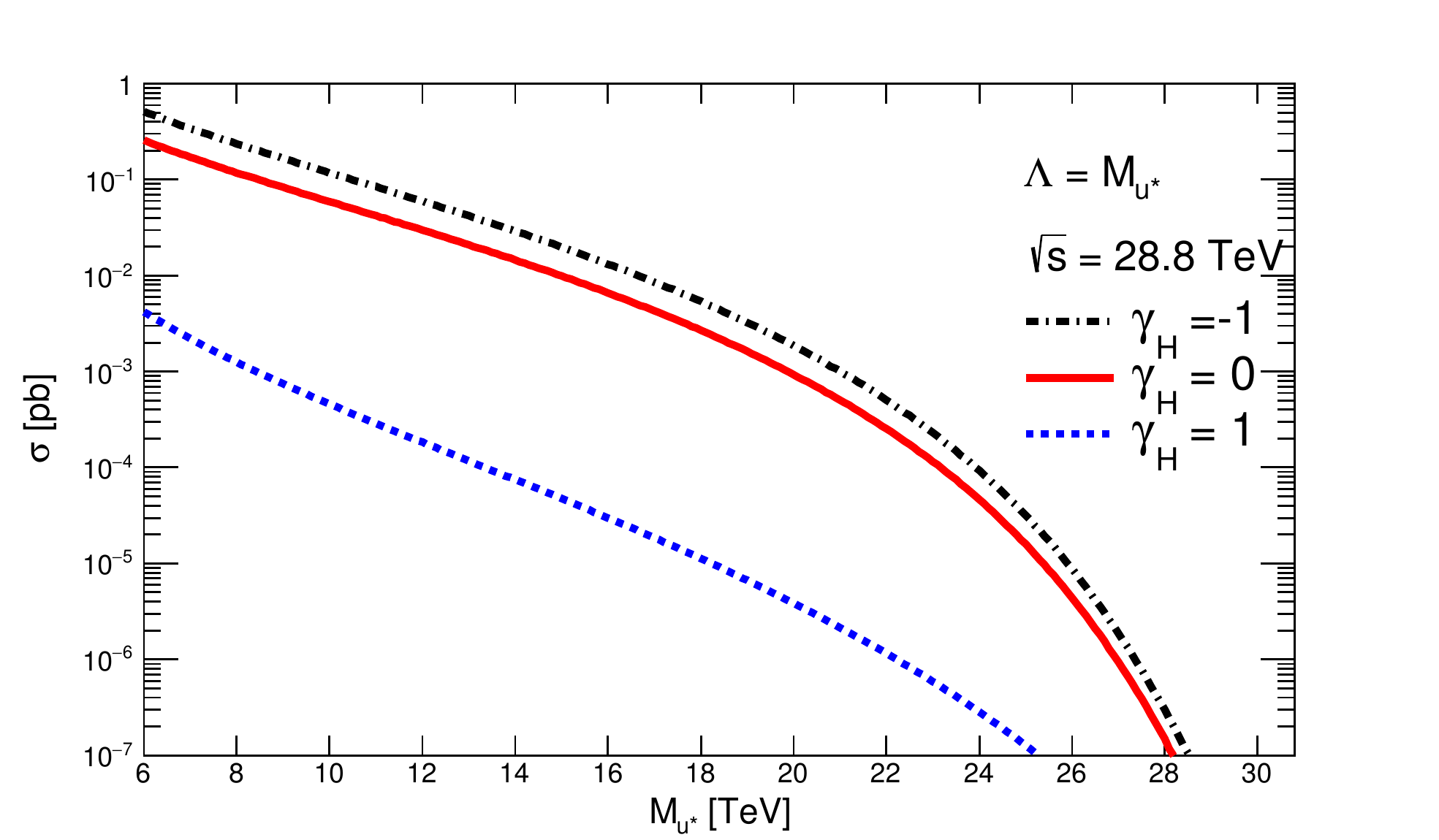}
	\caption{\label{fig:sc28800} $u^*$ cross section values  with respect to  its mass  for proton collision with both polarized and unpolarized photon beams at $\sqrt{s} = 28.8$ TeV.  On the left plot compositeness scale was chosen as 30 TeV and on the right plot $u^*$ mass was taken the same as compositeness scale.}
\end{figure}

Figure \ref{fig:sc28800} represents the same plots like the previous one but this time, the center of mass energy is 28.8 TeV. It is seen that excited quark production could be achieved at higher mass values than previous collider option due to high center of mass energy in this collider option. 

\subsection{\label{subsec:signal}Signal and Background Analysis} 
\subsubsection{\label{subsubsec:finalstate} Final State Distributions and Cut Determination}
Our signal process is  $\gamma + p \rightarrow u^* + X\rightarrow   u + g   +X$, therefore,  background processes are represented by  $\gamma + p \rightarrow  j + j + X$, where $j$ denotes   $u, \; \bar{u},\;d,\;\bar{d},\; c,\; \bar{c},\; s,\; \bar{s},\; b,$ $\;\bar{b}$ and $g$ jets. To assign cuts for identifying the signal from background, we examined at the both signal and background transverse momentum ($P_T$), the pseudo rapidity ($\eta$)  and the invariant mass distributions for the final state particles. Below, we present results for $\gamma_{\mathcal{H}}$ = -1 case which corresponds to maximal signal cross section values. It should be noted that we normalized cross section values to plot $P_T$ and $\eta$ distributions for obtaining the cuts. 

\begin{figure}[h!]
	\centering
	\includegraphics[width=0.5\columnwidth]{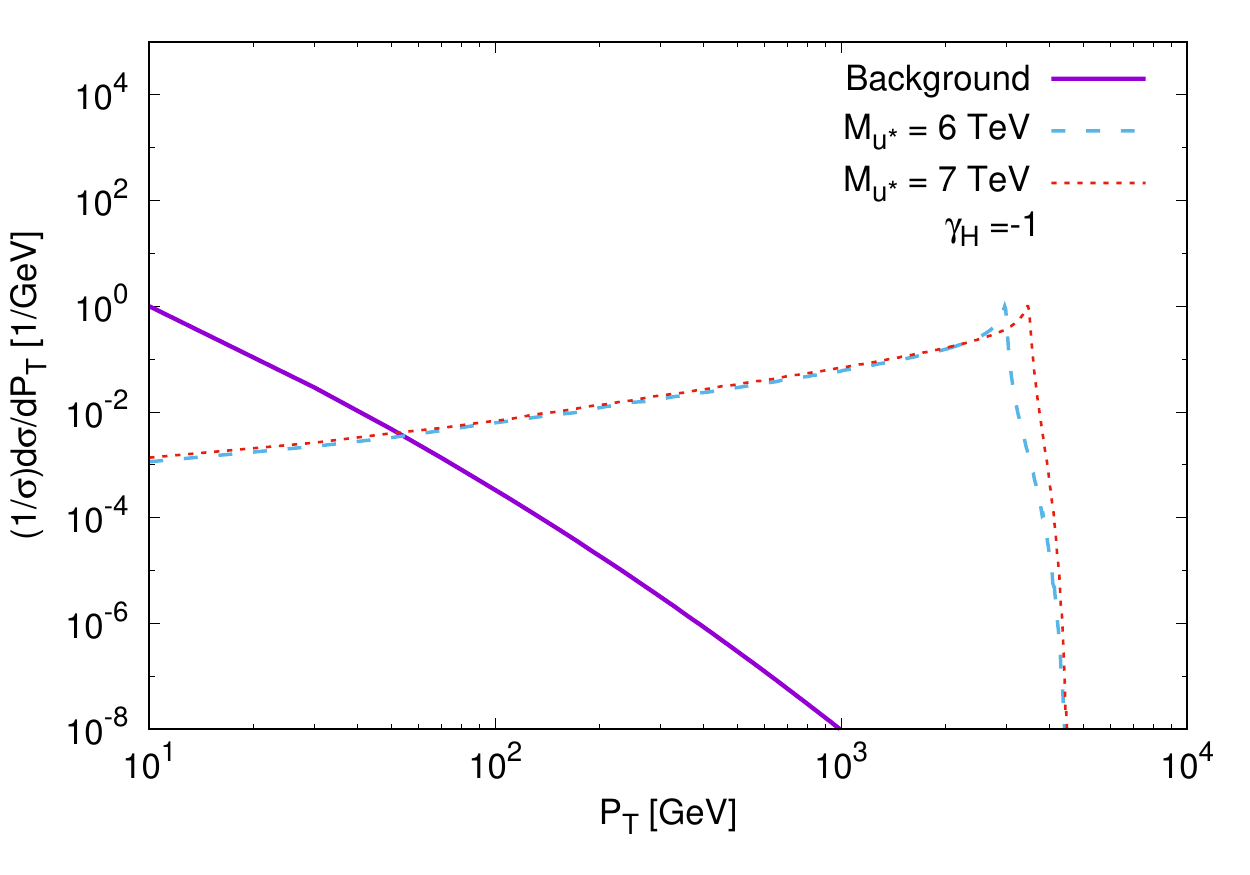}%
	\includegraphics[width=0.5\columnwidth]{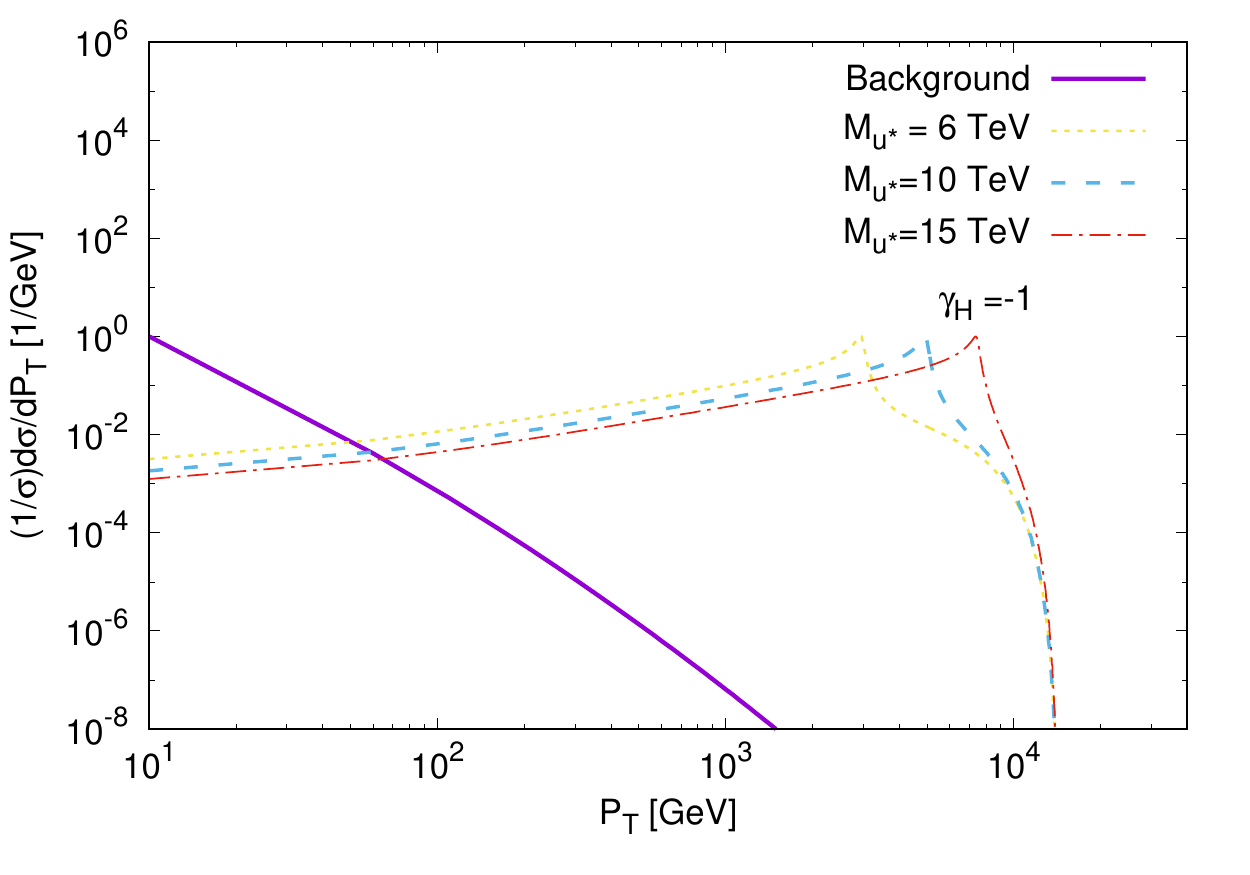}%
	\caption{\label{fig:CM1032Pt} Normalized $P_T$ distributions of background and signal processes for  $\sqrt{s} = 9.1$ TeV  at the left  panel and  for $\sqrt{s} = 28.8$ TeV at the right panel.}
\end{figure}

$P_T$ distributions of the signal are the same for the two final state particles ($u, g$) and the  background $P_T$ distributions’ final state particles that are jets, defined above. Figure \ref{fig:CM1032Pt} demonstrates  $P_T$ distributions of the signal  and the background  final state jets for both two center of mass energy options.  As expected for a single resonance, the usual  Jacobean peaks appear for all mass values (6, 7, 10 and 15 TeV) of the signal. It is seen that when the applied $P_T$ cut was taken 500 GeV for the $\sqrt{s} = 9.1$ TeV and 1000 GeV for the $\sqrt{s} = 28.8$ TeV, the background  was  reduced almost completely  but the signal was remained nearly unchanged. 

\begin{figure}[h!]
	\centering
	\includegraphics[width=0.5\columnwidth]{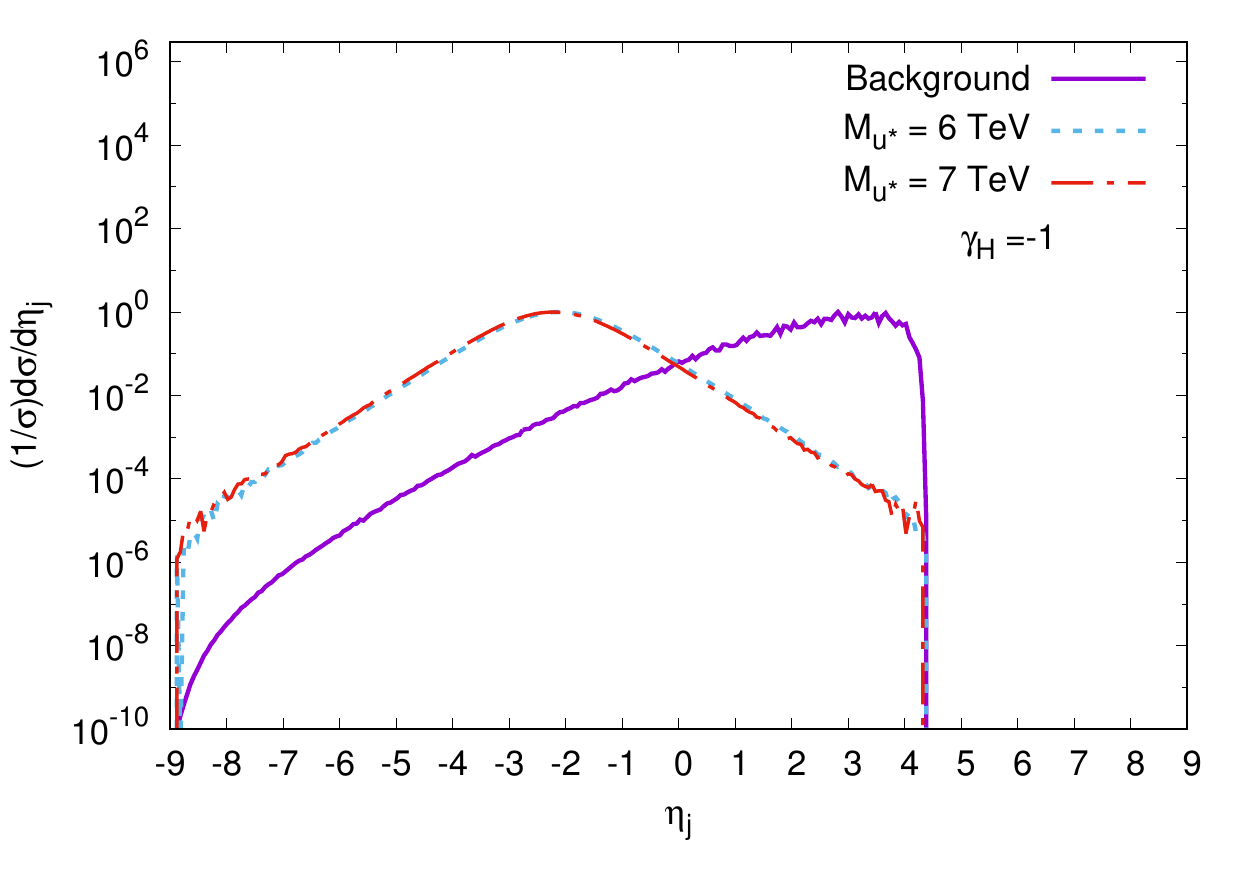}%
	\includegraphics[width=0.5\columnwidth]{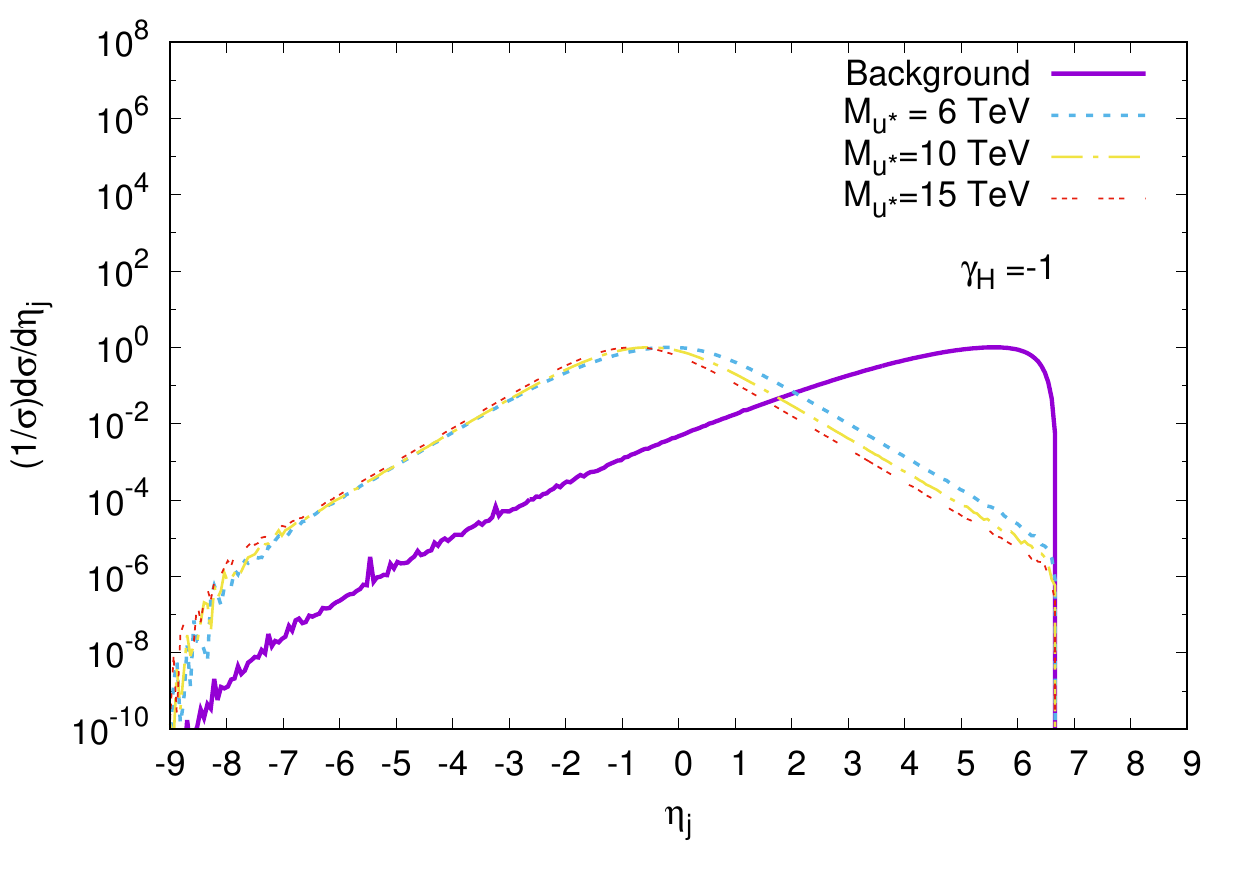}%
	\caption{\label{fig:CM10eta} Normalized  $\eta$ distributions of background and signal processes for $\sqrt{s} = 9.1$ TeV at the left panel and for $\sqrt{s} = 28.8$ TeV at the right panel.}
\end{figure}

When the colliding beams have different energies,  asymmetry occurs in signal and background distributions. So, we  extracted  $\eta$ cuts using  signal and  background final state jet distributions at their crossing point of their right side limits, that are shown for both center of mass energies in Figure \ref{fig:CM10eta}. $\eta$ distributions are presented as sum of both final state particles contributions because it is hard to identify gluon and u-quark apart.  On the other hand, we applied $\eta$ cuts as -5.2 for the left hand side of the $\eta$ distributions, this value was taken from the CMS experiment forward sub-detector limits \cite{cms2015}. We summarized all $\eta$ cuts in Table \ref{tab:etacut}.

\begin{figure}[h!]
	\centering
	\includegraphics[width=0.5\columnwidth]{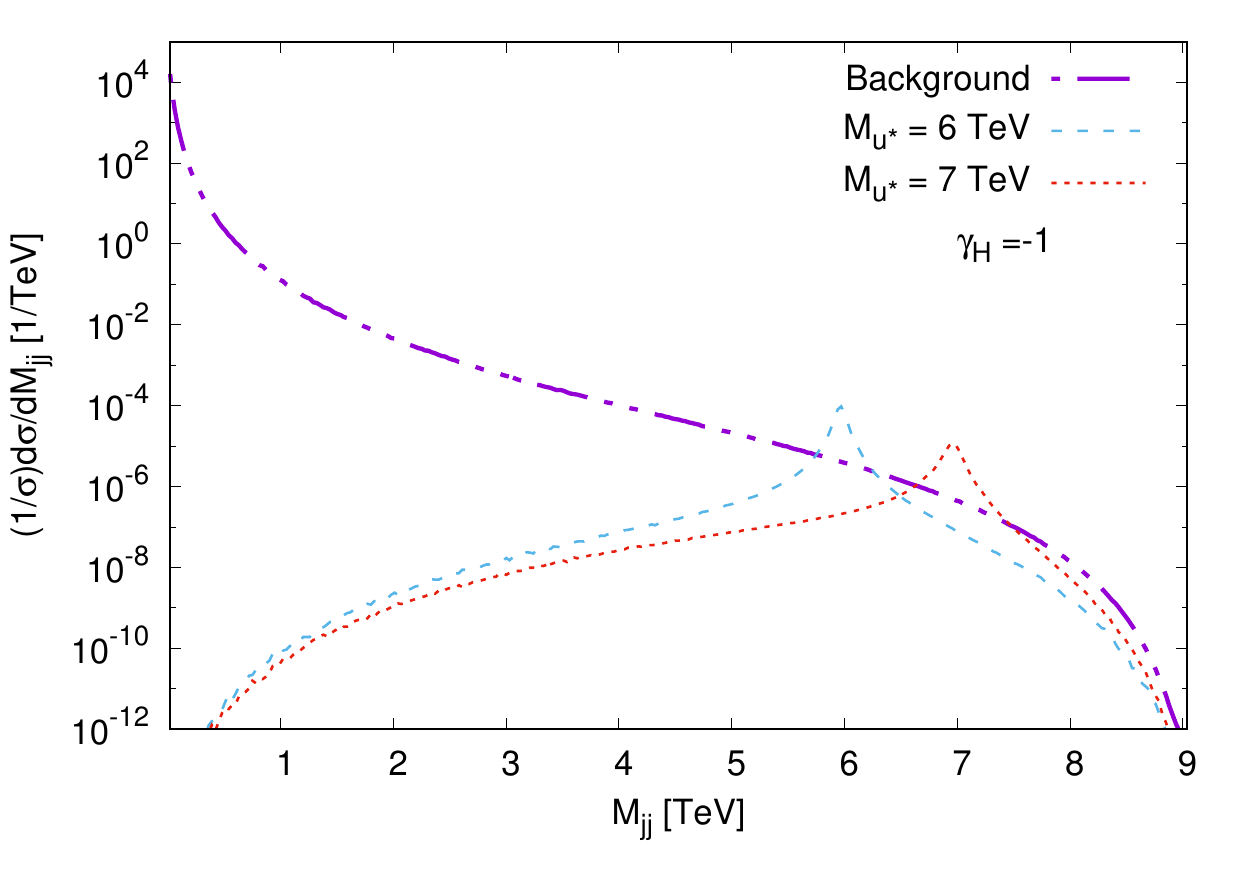}%
	\includegraphics[width=0.5\columnwidth]{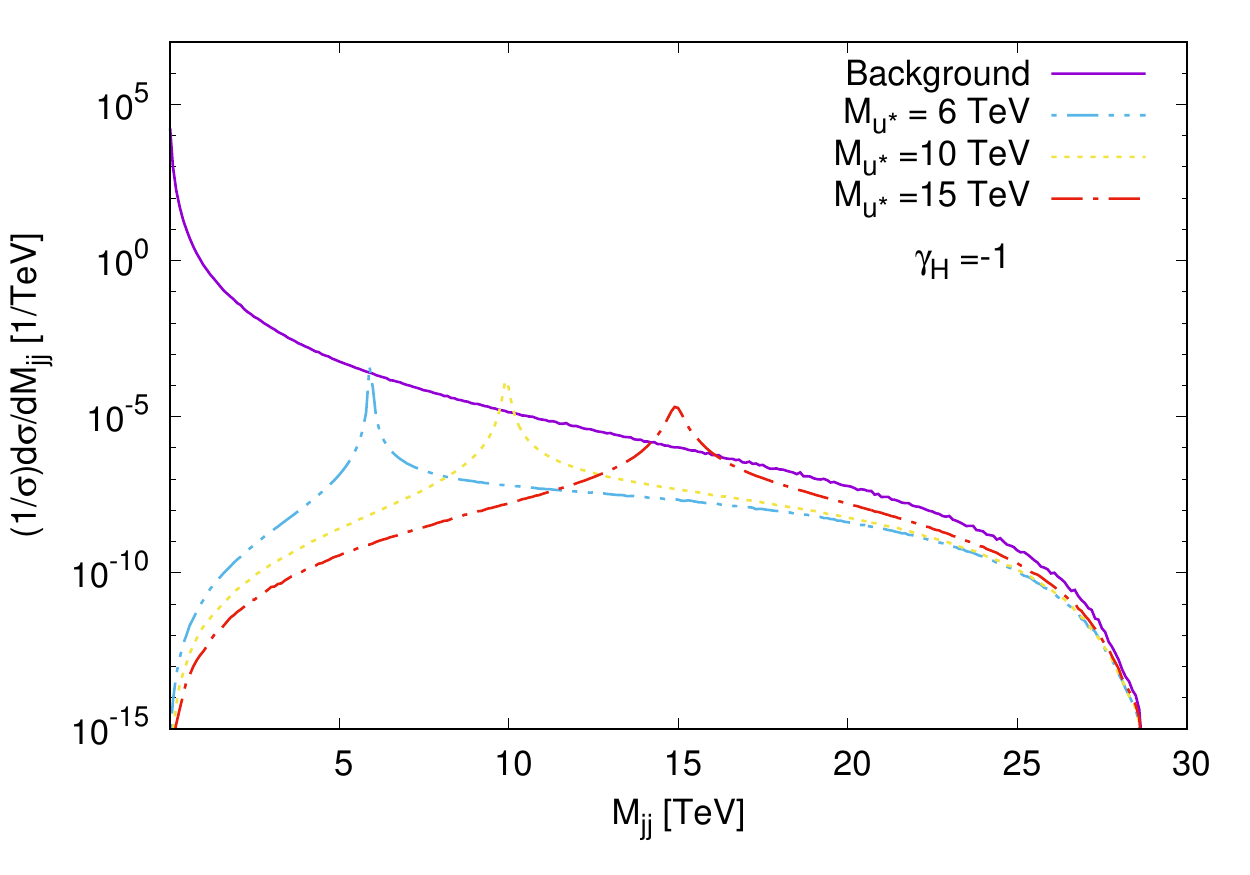}%
	\caption{\label{fig:CM1032inv} Signal and background  invariant mass distributions for  $\sqrt{s} = 9.1$ TeV (the left panel) and $\sqrt{s} = 28.8$ TeV (the right panel) with $\gamma_{\mathcal{H}}= -1$.}
\end{figure}

Invariant mass distributions for signal and background  processes are presented in Figure \ref{fig:CM1032inv}. It is seen that  signal peak values are above the background, so we determined invariant mass cut as $M_{u^*}-2\Gamma_{u^*}$ and $M_{u^*}+2\Gamma_{u^*}$ mass window, where  $M_{u^*}$ is  $u^*$ mass and $\Gamma_{u^*}$ is the decay widths of the $u^*$. 

\begin{table*} [h!]
	\caption{\label{tab:etacut} List of the pseudo rapidity cut limits for both center of mass energy options.}
	\begin{tabular*}{\textwidth}{@{\extracolsep{\fill}}c|cc|cc|cc|cc@{}} \hline
		$\sqrt{s}$ (TeV) & \multicolumn{4}{c|}{9.1 } & \multicolumn{4}{c}{28.8 }   \\ \hline 
		$\gamma_{\mathcal{H}}$  &   \multicolumn{2}{c|}{-1} & \multicolumn{2}{c|}{0} &  \multicolumn{2}{c|}{-1} &\multicolumn{2}{c}{0} \\ \hline
		Cut Limits   & Min& Max& Min& Max& Min& Max& Min& Max  \\ \hline
		$\eta_{j}$	& -5.2 & 0.0 & -5.2 & -0.2 & -5.2 &  2.1& -5.2 & 2.0\\  \hline	
	\end{tabular*}
\end{table*}

\subsubsection{Mass Limits Dependence on Integrated Luminosity and Photon Beam Polarization}
To extract the signal from the background, we used the cuts that were determined by distribution plots in the previous subsection. After that, Equation \ref{eq:significance} was utilized  to calculate the statistical significance,

\begin{equation}
\label{eq:significance}
S = \frac{\sigma_{s}}{\sqrt{\sigma_{s}+\sigma_{B}}}  \sqrt{\mathcal{L}_{int}}
\end{equation}     

where, $\sigma_{s}$ and $\sigma_{B}$ are the signal and background cross section values, respectively  and $\mathcal{L}_{int}$ is the integrated luminosity.  Obtained $u^*$ mass limits were listed in Tables \ref{tab:lumMass10} and \ref{tab:lumMass32} for both center of mass energies 9.1 TeV and 28.8 TeV colliders, respectively. According to Table \ref{tab:lcpar},  integrated luminosity values are 10-100 $fb^{-1}$ for ILC$\otimes$FCC and 1-10 $fb^{-1}$ for PWFA-LC$\otimes$FCC options. As expected, higher integrated luminosity increased mass limits for $u^*$. Besides, it can be seen from Tables \ref{tab:lumMass10} and \ref{tab:lumMass32} that photon beam polarization  enhanced $u^*$ mass limits 0.21 TeV for 9.1 TeV CM and approximately  1.5 TeV for 28.8 TeV CM at their upper luminosity values if compared to unpolarized photon beam-proton collisions. In addition, the attainable best $u^*$ mass limits could be achieved when the $\Lambda = M_{u^*}$. 

\begin{table*}[h!]
	\centering
	\caption{Excited u quark mass limits for 9.1 TeV center of mass energy $\gamma p$ collider.}
	\label{tab:lumMass10}
	
	\begin{tabular*}{\textwidth}{@{\extracolsep{\fill}}ll|l|c|c|c|c|c|c|c@{}}
		\hline
		\multicolumn{2}{c|}{$\sqrt{s}$ }  & \multicolumn{8}{c}{9.1 TeV }                                                                                      \\ \hline
		\multicolumn{2}{c|}{$\mathcal{L}_{int}$ }             & \multicolumn{4}{c|}{10 $fb^{-1}$}                                   & \multicolumn{4}{c}{100 $fb^{-1}$}                                  \\ \hline
		\multicolumn{2}{c|}{$\Lambda$}    & \multicolumn{2}{c|}{10 TeV} & \multicolumn{2}{c|}{$M_{u^*}$} & \multicolumn{2}{c|}{10 TeV} & \multicolumn{2}{c}{$M_{u^*}$} \\ \hline
		\multicolumn{2}{c|}{$\gamma_{\mathcal{H}}$}               & -1      & 0         & -1      & 0           & -1      & 0        & -1      &  0           \\ \hline
		\multirow{3}{*}{\begin{tabular}[c]{@{}c@{}}Mass Limits\\ (TeV)\end{tabular}} & 5$\sigma$ & 6.97 & 6.58   & 7.27  & 6.96  & 7.82  & 7.60  & 7.99  &7.78  \\ \cline{2-10}    
		& 3$\sigma$ & 7.41  & 7.11  & 7.62  & 7.37   &8.08  & 7.90  &  8.23  & 8.05    \\ \cline{2-10} 
		& 2$\sigma$ & 7.68 & 7.43 &  7.86 & 7.64 & 8.24 &8.10  & 8.40 &  8.24   \\  \hline
	\end{tabular*}
	
\end{table*}

\begin{table*}[h!]
	\centering
	\caption{Excited u quark mass limits for 28.8 TeV center of mass energy $\gamma p$ collider.}
	\label{tab:lumMass32}
	
	\begin{tabular*}{\textwidth}{@{\extracolsep{\fill}}ll|c|c|c|c|c|c|c|c@{}}
		\hline
		\multicolumn{2}{c|}{$\sqrt{s}$ }  & \multicolumn{8}{c}{28.8 TeV }                                                                                      \\ \hline
		\multicolumn{2}{c|}{$\mathcal{L}_{int}$ }             & \multicolumn{4}{c|}{1 $fb^{-1}$}                                   & \multicolumn{4}{c}{10 $fb^{-1}$}                                  \\ \hline
		\multicolumn{2}{c|}{$\Lambda$}    & \multicolumn{2}{c|}{15 TeV} & \multicolumn{2}{c|}{$M_{u^*}$} & \multicolumn{2}{c|}{30 TeV} & \multicolumn{2}{c}{$M_{u^*}$} \\ \hline
		\multicolumn{2}{c|}{$\gamma_{\mathcal{H}}$}               & -1      & 0     & -1      & 0    & -1      & 0       & -1      & 0       \\ \hline
		\multirow{3}{*}{\begin{tabular}[c]{@{}c@{}}Mass Limits\\ (TeV)\end{tabular}} & 5$\sigma$ &13.8  & 8.94    & 14.2    & 12.1 &  17.1  & 14.5  & 19.4    & 17.9    \\ \cline{2-10} 
		& 3$\sigma$ & 17.4   & 14.9  & 16.8  & 14.9    &19.7 & 17.9   & 21.1    & 19.9   \\ \cline{2-10} 
		& 2$\sigma$ & 19.5 & 17.6  & 18.5  & 16.9  & 21.3    & 19.8  & 22.2   & 21.2 \\ \hline
	\end{tabular*}
	
\end{table*}

\begin{figure}[h!]
	\centering
	\includegraphics[width=0.5\columnwidth]{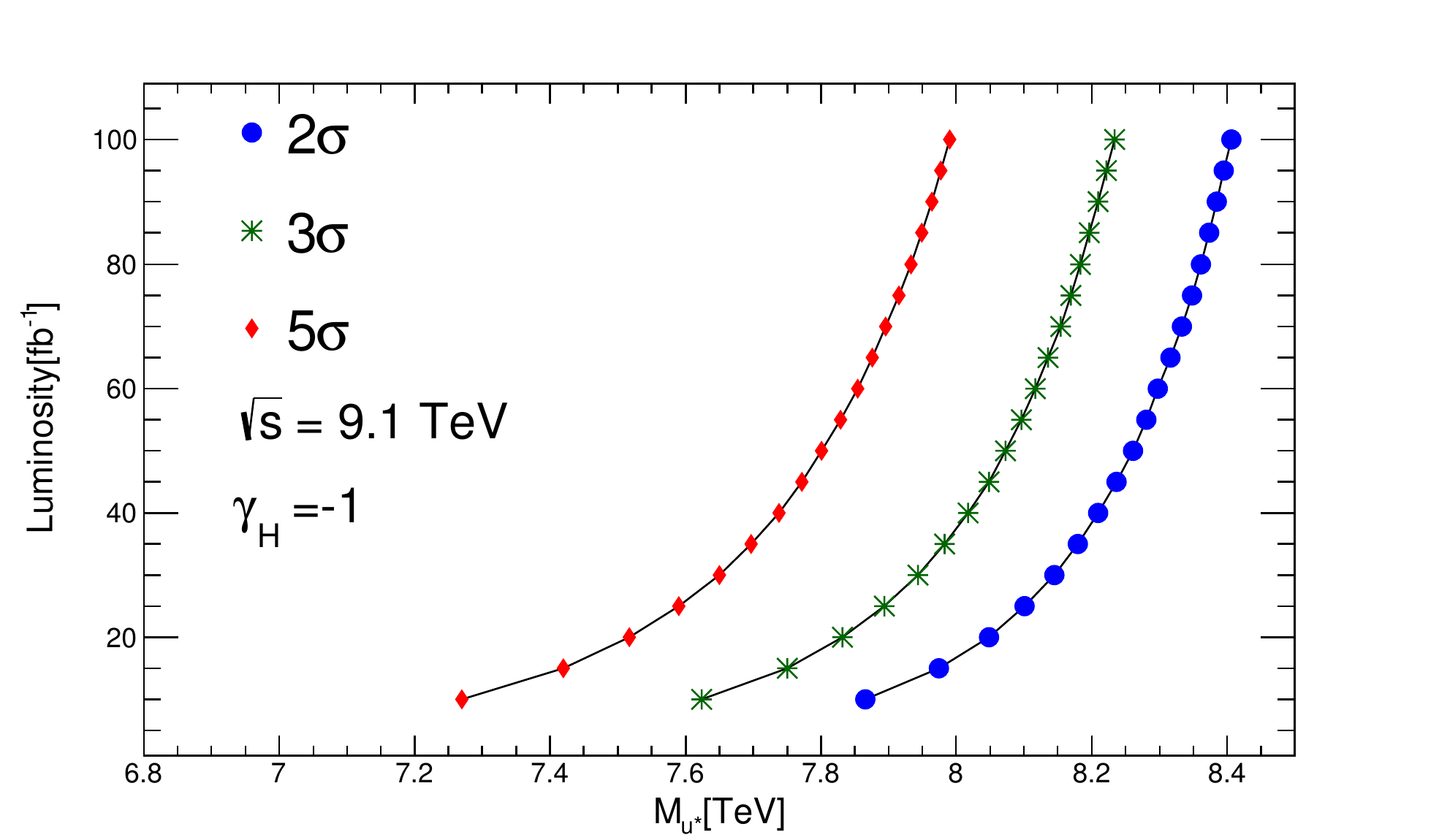}%
	\includegraphics[width=0.5\columnwidth]{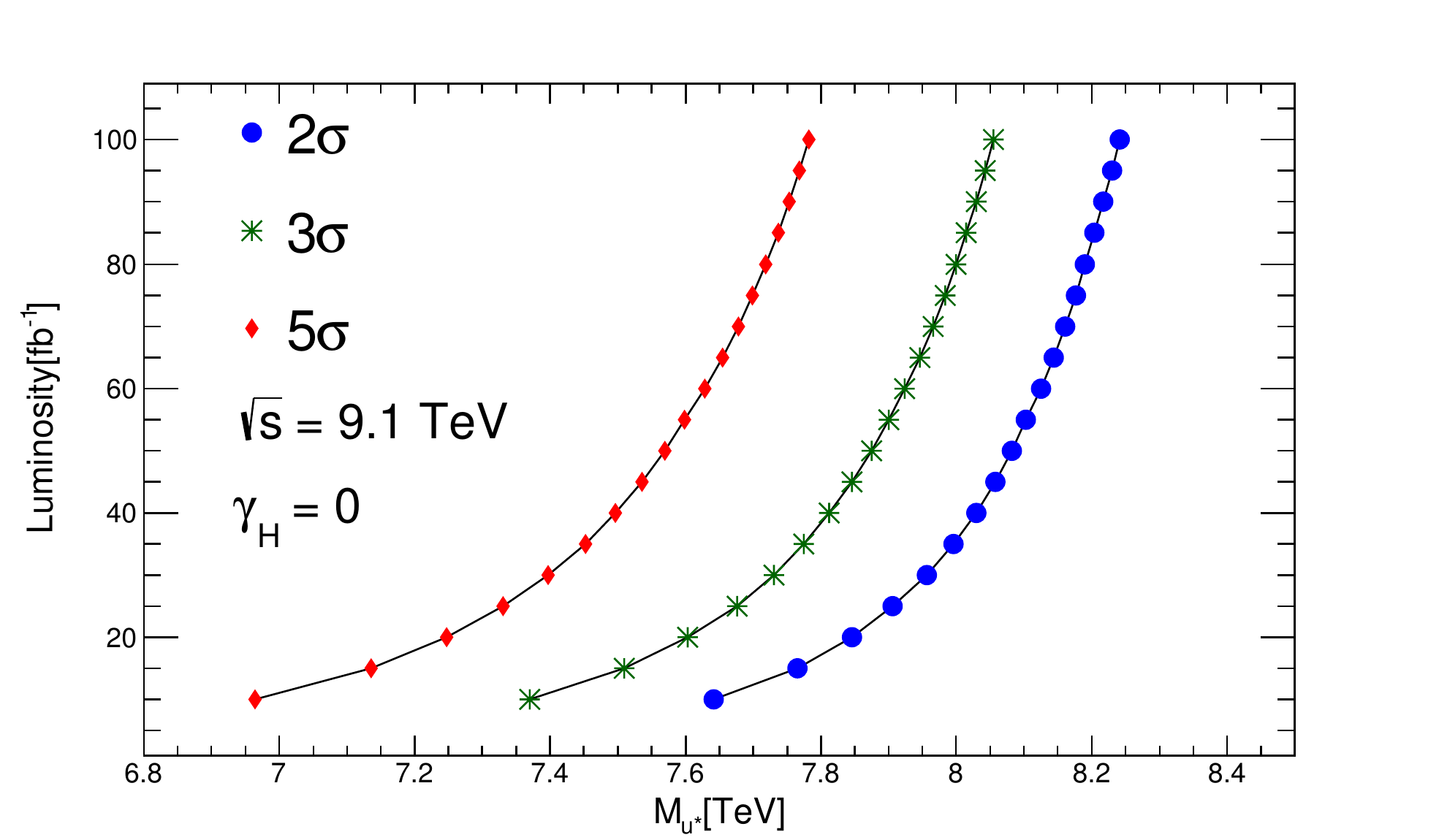}\\%
	\includegraphics[width=0.5\columnwidth]{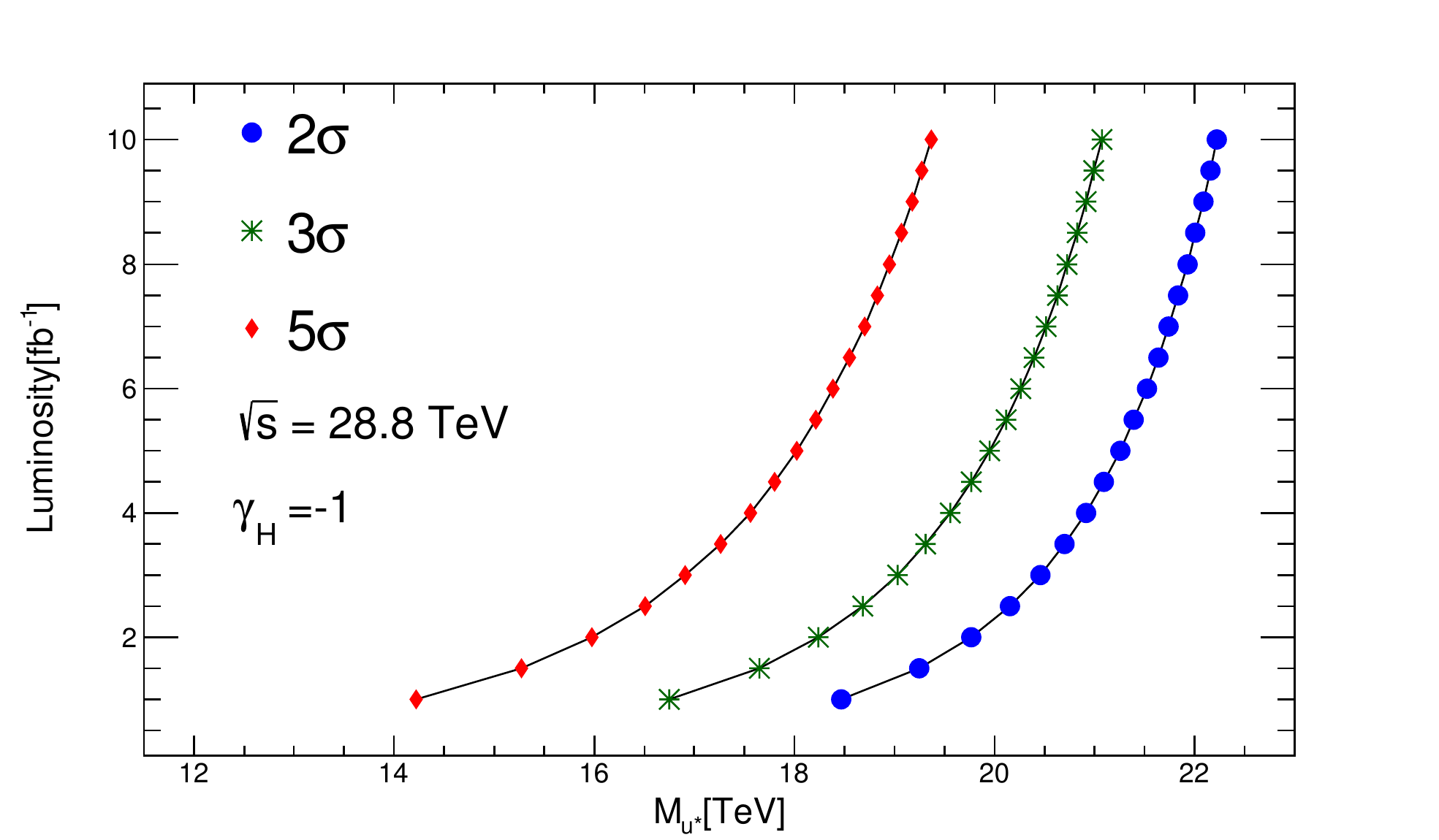}%
	\includegraphics[width=0.5\columnwidth]{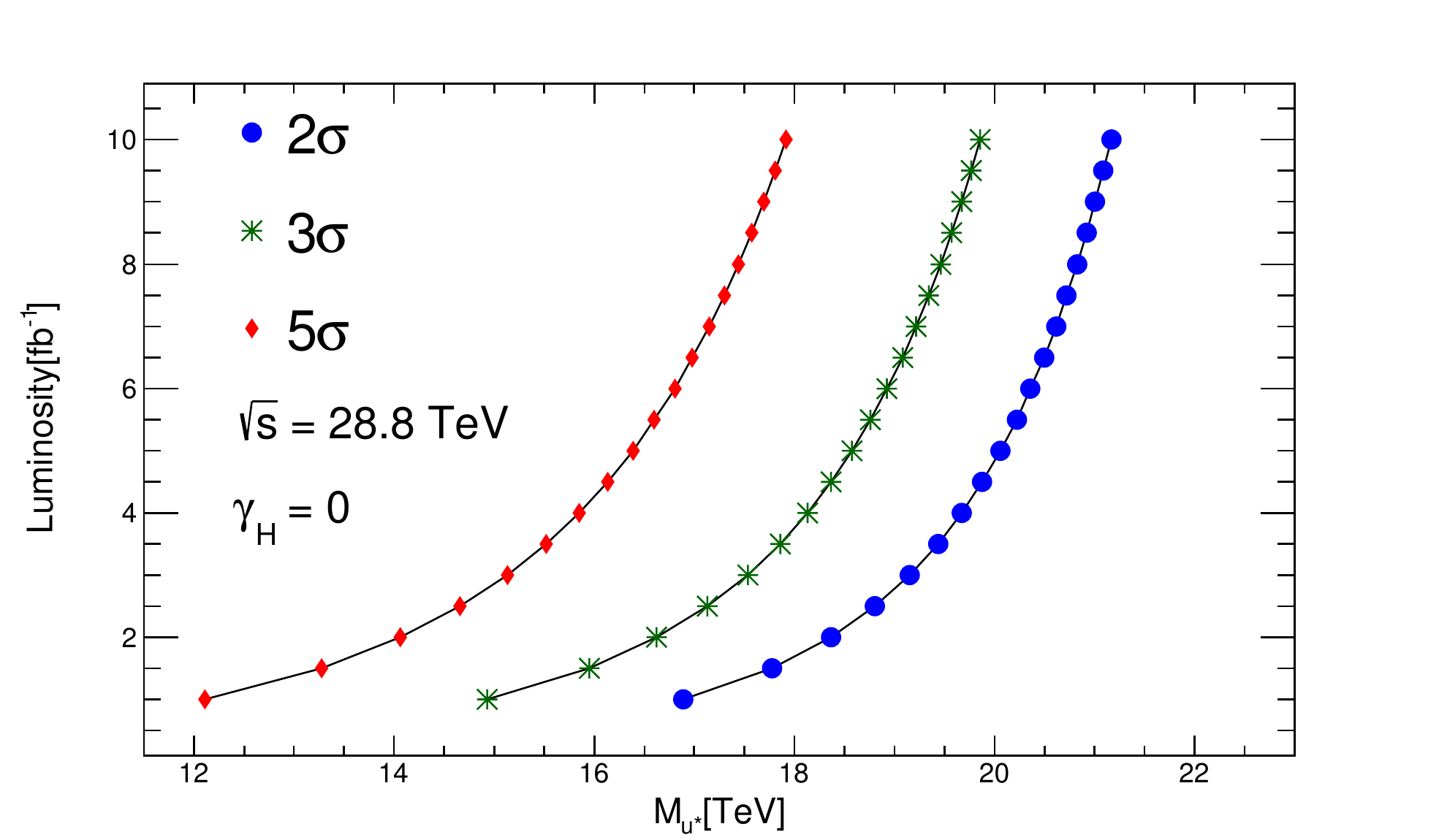}%
	\caption{\label{fig:lumLamMass} The first row represents luminosity and $u^*$ mass relations for  $\sqrt{s} = 9.1$ TeV and the second row shows the same relations for $\sqrt{s} = 28.8$ TeV with $\Lambda = M_{u^*}$  at three different significance values. The left column corresponds to $\gamma_{\mathcal{H}}=-1$  and the right panel corresponds to  $\gamma_{\mathcal{H} }$= 0.}
\end{figure}

In Figure \ref{fig:lumLamMass}, we scanned luminosity values needed for the discovery ($5\sigma$), observation ($3\sigma$) and exclusion ($2\sigma$) of $u^*$ as a function of its mass. It is seen that photon beam polarization enhanced  attainable mass limits of $u^*$.

\subsubsection{Attainable Compositeness Scale}

\begin{figure}[h!]
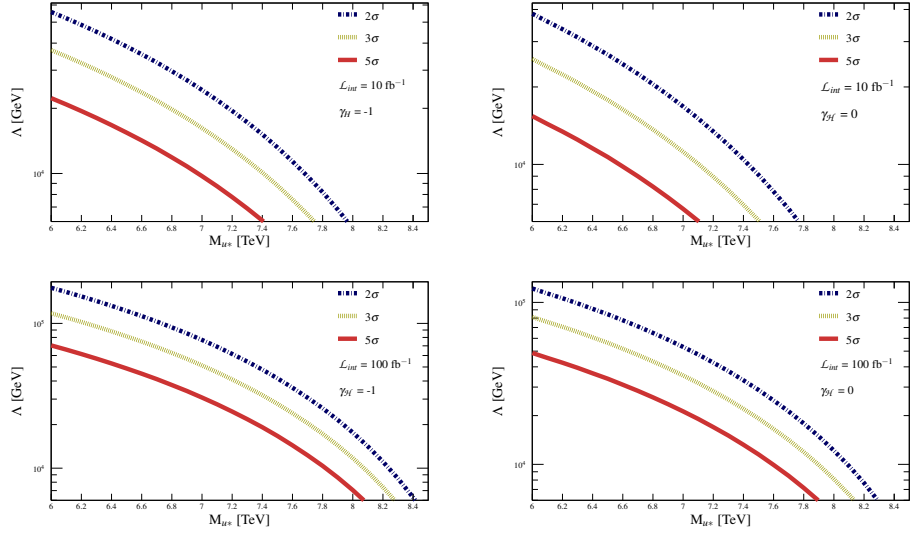

	\centering
	\scalebox{0.31}{\input{sigAllCan_Lumi10_LHPol.tex}}
	\scalebox{0.31}{\begin{tikzpicture}
\pgfdeclareplotmark{cross} {
\pgfpathmoveto{\pgfpoint{-0.3\pgfplotmarksize}{\pgfplotmarksize}}
\pgfpathlineto{\pgfpoint{+0.3\pgfplotmarksize}{\pgfplotmarksize}}
\pgfpathlineto{\pgfpoint{+0.3\pgfplotmarksize}{0.3\pgfplotmarksize}}
\pgfpathlineto{\pgfpoint{+1\pgfplotmarksize}{0.3\pgfplotmarksize}}
\pgfpathlineto{\pgfpoint{+1\pgfplotmarksize}{-0.3\pgfplotmarksize}}
\pgfpathlineto{\pgfpoint{+0.3\pgfplotmarksize}{-0.3\pgfplotmarksize}}
\pgfpathlineto{\pgfpoint{+0.3\pgfplotmarksize}{-1.\pgfplotmarksize}}
\pgfpathlineto{\pgfpoint{-0.3\pgfplotmarksize}{-1.\pgfplotmarksize}}
\pgfpathlineto{\pgfpoint{-0.3\pgfplotmarksize}{-0.3\pgfplotmarksize}}
\pgfpathlineto{\pgfpoint{-1.\pgfplotmarksize}{-0.3\pgfplotmarksize}}
\pgfpathlineto{\pgfpoint{-1.\pgfplotmarksize}{0.3\pgfplotmarksize}}
\pgfpathlineto{\pgfpoint{-0.3\pgfplotmarksize}{0.3\pgfplotmarksize}}
\pgfpathclose
\pgfusepathqstroke
}
\pgfdeclareplotmark{cross*} {
\pgfpathmoveto{\pgfpoint{-0.3\pgfplotmarksize}{\pgfplotmarksize}}
\pgfpathlineto{\pgfpoint{+0.3\pgfplotmarksize}{\pgfplotmarksize}}
\pgfpathlineto{\pgfpoint{+0.3\pgfplotmarksize}{0.3\pgfplotmarksize}}
\pgfpathlineto{\pgfpoint{+1\pgfplotmarksize}{0.3\pgfplotmarksize}}
\pgfpathlineto{\pgfpoint{+1\pgfplotmarksize}{-0.3\pgfplotmarksize}}
\pgfpathlineto{\pgfpoint{+0.3\pgfplotmarksize}{-0.3\pgfplotmarksize}}
\pgfpathlineto{\pgfpoint{+0.3\pgfplotmarksize}{-1.\pgfplotmarksize}}
\pgfpathlineto{\pgfpoint{-0.3\pgfplotmarksize}{-1.\pgfplotmarksize}}
\pgfpathlineto{\pgfpoint{-0.3\pgfplotmarksize}{-0.3\pgfplotmarksize}}
\pgfpathlineto{\pgfpoint{-1.\pgfplotmarksize}{-0.3\pgfplotmarksize}}
\pgfpathlineto{\pgfpoint{-1.\pgfplotmarksize}{0.3\pgfplotmarksize}}
\pgfpathlineto{\pgfpoint{-0.3\pgfplotmarksize}{0.3\pgfplotmarksize}}
\pgfpathclose
\pgfusepathqfillstroke
}
\pgfdeclareplotmark{newstar} {
\pgfpathmoveto{\pgfqpoint{0pt}{\pgfplotmarksize}}
\pgfpathlineto{\pgfqpointpolar{44}{0.5\pgfplotmarksize}}
\pgfpathlineto{\pgfqpointpolar{18}{\pgfplotmarksize}}
\pgfpathlineto{\pgfqpointpolar{-20}{0.5\pgfplotmarksize}}
\pgfpathlineto{\pgfqpointpolar{-54}{\pgfplotmarksize}}
\pgfpathlineto{\pgfqpointpolar{-90}{0.5\pgfplotmarksize}}
\pgfpathlineto{\pgfqpointpolar{234}{\pgfplotmarksize}}
\pgfpathlineto{\pgfqpointpolar{198}{0.5\pgfplotmarksize}}
\pgfpathlineto{\pgfqpointpolar{162}{\pgfplotmarksize}}
\pgfpathlineto{\pgfqpointpolar{134}{0.5\pgfplotmarksize}}
\pgfpathclose
\pgfusepathqstroke
}
\pgfdeclareplotmark{newstar*} {
\pgfpathmoveto{\pgfqpoint{0pt}{\pgfplotmarksize}}
\pgfpathlineto{\pgfqpointpolar{44}{0.5\pgfplotmarksize}}
\pgfpathlineto{\pgfqpointpolar{18}{\pgfplotmarksize}}
\pgfpathlineto{\pgfqpointpolar{-20}{0.5\pgfplotmarksize}}
\pgfpathlineto{\pgfqpointpolar{-54}{\pgfplotmarksize}}
\pgfpathlineto{\pgfqpointpolar{-90}{0.5\pgfplotmarksize}}
\pgfpathlineto{\pgfqpointpolar{234}{\pgfplotmarksize}}
\pgfpathlineto{\pgfqpointpolar{198}{0.5\pgfplotmarksize}}
\pgfpathlineto{\pgfqpointpolar{162}{\pgfplotmarksize}}
\pgfpathlineto{\pgfqpointpolar{134}{0.5\pgfplotmarksize}}
\pgfpathclose
\pgfusepathqfillstroke
}
\definecolor{c}{rgb}{1,1,1};
\draw [color=c, fill=c] (0,0) rectangle (20,11.6806);
\draw [color=c, fill=c] (2,1.16806) rectangle (18,10.5125);
\definecolor{c}{rgb}{0,0,0};
\draw [c,line width=0.9] (2,1.16806) -- (2,10.5125) -- (18,10.5125) -- (18,1.16806) -- (2,1.16806);
\definecolor{c}{rgb}{1,1,1};
\draw [color=c, fill=c] (2,1.16806) rectangle (18,10.5125);
\definecolor{c}{rgb}{0,0,0};
\draw [c,line width=0.9] (2,1.16806) -- (2,10.5125) -- (18,10.5125) -- (18,1.16806) -- (2,1.16806);
\draw [c,line width=0.9] (2,1.16806) -- (18,1.16806);
\draw (10,0.327056) node[scale=1.56475, color=c, rotate=0]{M$_{u*}$ [TeV]};
\draw [c,line width=0.9] (2,1.44839) -- (2,1.16806);
\draw [c,line width=0.9] (3.28,1.44839) -- (3.28,1.16806);
\draw [c,line width=0.9] (4.56,1.44839) -- (4.56,1.16806);
\draw [c,line width=0.9] (5.84,1.44839) -- (5.84,1.16806);
\draw [c,line width=0.9] (7.12,1.44839) -- (7.12,1.16806);
\draw [c,line width=0.9] (8.4,1.44839) -- (8.4,1.16806);
\draw [c,line width=0.9] (9.68,1.44839) -- (9.68,1.16806);
\draw [c,line width=0.9] (10.96,1.44839) -- (10.96,1.16806);
\draw [c,line width=0.9] (12.24,1.44839) -- (12.24,1.16806);
\draw [c,line width=0.9] (13.52,1.44839) -- (13.52,1.16806);
\draw [c,line width=0.9] (14.8,1.44839) -- (14.8,1.16806);
\draw [c,line width=0.9] (16.08,1.44839) -- (16.08,1.16806);
\draw [c,line width=0.9] (17.36,1.44839) -- (17.36,1.16806);
\draw [c,line width=0.9] (17.36,1.44839) -- (17.36,1.16806);
\draw [anchor=base] (2,0.782598) node[scale=0.900036, color=c, rotate=0]{6};
\draw [anchor=base] (3.28,0.782598) node[scale=0.900036, color=c, rotate=0]{6.2};
\draw [anchor=base] (4.56,0.782598) node[scale=0.900036, color=c, rotate=0]{6.4};
\draw [anchor=base] (5.84,0.782598) node[scale=0.900036, color=c, rotate=0]{6.6};
\draw [anchor=base] (7.12,0.782598) node[scale=0.900036, color=c, rotate=0]{6.8};
\draw [anchor=base] (8.4,0.782598) node[scale=0.900036, color=c, rotate=0]{7};
\draw [anchor=base] (9.68,0.782598) node[scale=0.900036, color=c, rotate=0]{7.2};
\draw [anchor=base] (10.96,0.782598) node[scale=0.900036, color=c, rotate=0]{7.4};
\draw [anchor=base] (12.24,0.782598) node[scale=0.900036, color=c, rotate=0]{7.6};
\draw [anchor=base] (13.52,0.782598) node[scale=0.900036, color=c, rotate=0]{7.8};
\draw [anchor=base] (14.8,0.782598) node[scale=0.900036, color=c, rotate=0]{8};
\draw [anchor=base] (16.08,0.782598) node[scale=0.900036, color=c, rotate=0]{8.2};
\draw [anchor=base] (17.36,0.782598) node[scale=0.900036, color=c, rotate=0]{8.4};
\draw [c,line width=0.9] (2,10.5125) -- (18,10.5125);
\draw [c,line width=0.9] (2,10.2322) -- (2,10.5125);
\draw [c,line width=0.9] (3.28,10.2322) -- (3.28,10.5125);
\draw [c,line width=0.9] (4.56,10.2322) -- (4.56,10.5125);
\draw [c,line width=0.9] (5.84,10.2322) -- (5.84,10.5125);
\draw [c,line width=0.9] (7.12,10.2322) -- (7.12,10.5125);
\draw [c,line width=0.9] (8.4,10.2322) -- (8.4,10.5125);
\draw [c,line width=0.9] (9.68,10.2322) -- (9.68,10.5125);
\draw [c,line width=0.9] (10.96,10.2322) -- (10.96,10.5125);
\draw [c,line width=0.9] (12.24,10.2322) -- (12.24,10.5125);
\draw [c,line width=0.9] (13.52,10.2322) -- (13.52,10.5125);
\draw [c,line width=0.9] (14.8,10.2322) -- (14.8,10.5125);
\draw [c,line width=0.9] (16.08,10.2322) -- (16.08,10.5125);
\draw [c,line width=0.9] (17.36,10.2322) -- (17.36,10.5125);
\draw [c,line width=0.9] (17.36,10.2322) -- (17.36,10.5125);
\draw [c,line width=0.9] (2,1.16806) -- (2,10.5125);
\draw (0.56,5.84029) node[scale=1.56475, color=c, rotate=90]{$\Lambda$ [GeV]};
\draw [c,line width=0.9] (2.24,1.16884) -- (2,1.16884);
\draw [c,line width=0.9] (2.24,1.90573) -- (2,1.90573);
\draw [c,line width=0.9] (2.24,2.54404) -- (2,2.54404);
\draw [c,line width=0.9] (2.24,3.10708) -- (2,3.10708);
\draw [c,line width=0.9] (2.48,3.61073) -- (2,3.61073);
\draw [anchor= east] (1.844,3.61073) node[scale=0.900036, color=c, rotate=0]{$10^{4}$};
\draw [c,line width=0.9] (2.24,6.92417) -- (2,6.92417);
\draw [c,line width=0.9] (2.24,8.8624) -- (2,8.8624);
\draw [c,line width=0.9] (2.24,10.2376) -- (2,10.2376);
\draw [c,line width=0.9] (18,1.16806) -- (18,10.5125);
\draw [c,line width=0.9] (17.76,1.16884) -- (18,1.16884);
\draw [c,line width=0.9] (17.76,1.90573) -- (18,1.90573);
\draw [c,line width=0.9] (17.76,2.54404) -- (18,2.54404);
\draw [c,line width=0.9] (17.76,3.10708) -- (18,3.10708);
\draw [c,line width=0.9] (17.52,3.61073) -- (18,3.61073);
\draw [c,line width=0.9] (17.76,6.92417) -- (18,6.92417);
\draw [c,line width=0.9] (17.76,8.8624) -- (18,8.8624);
\draw [c,line width=0.9] (17.76,10.2376) -- (18,10.2376);
\definecolor{c}{rgb}{0,0,0.4};
\draw [c,dash pattern=on 4.00pt off 2.40pt on 0.80pt off 2.40pt ,line width=5.4] (2,10.0572) -- (2.64,9.72056) -- (3.28,9.39501) -- (3.92,9.04836) -- (4.56,8.6749) -- (5.2,8.30749) -- (5.84,7.89846) -- (6.48,7.48325) -- (7.12,7.04032) --
 (7.76,6.57701) -- (8.4,6.08827) -- (9.04,5.57516) -- (9.68,5.02083) -- (10.32,4.44662) -- (10.96,3.82106) -- (11.6,3.16561) -- (12.24,2.45117) -- (12.88,1.70732) -- (13.3007,1.16806);
\definecolor{c}{rgb}{0.6,0.6,0};
\draw [c,dash pattern=on 0.80pt off 1.60pt ,line width=5.4] (2,8.11894) -- (2.64,7.78234) -- (3.28,7.45677) -- (3.92,7.11012) -- (4.56,6.73667) -- (5.2,6.36926) -- (5.84,5.96023) -- (6.48,5.54502) -- (7.12,5.1021) -- (7.76,4.63876) -- (8.4,4.15005)
 -- (9.04,3.63691) -- (9.68,3.08258) -- (10.32,2.50839) -- (10.96,1.88281) -- (11.6,1.22737) -- (11.6531,1.16806);
\definecolor{c}{rgb}{0.8,0.2,0.2};
\draw [c,line width=5.4] (2,5.67707) -- (2.64,5.34044) -- (3.28,5.01488) -- (3.92,4.66824) -- (4.56,4.29478) -- (5.2,3.92737) -- (5.84,3.51833) -- (6.48,3.10314) -- (7.12,2.6602) -- (7.76,2.19688) -- (8.4,1.70816) -- (9.04,1.19503) --
 (9.07114,1.16806);
\definecolor{c}{rgb}{0,0,0};
\draw (10,11.301) node[scale=1.2177, color=c, rotate=0]{ };
\draw [anchor=base west] (15.15,9.80681) node[scale=1.29711, color=c, rotate=0]{$2\sigma$};
\definecolor{c}{rgb}{0,0,0.4};
\draw [c,dash pattern=on 4.00pt off 2.40pt on 0.80pt off 2.40pt ,line width=5.4] (14.1725,10.0258) -- (14.9775,10.0258);
\definecolor{c}{rgb}{0,0,0};
\draw [anchor=base west] (15.15,8.83343) node[scale=1.29711, color=c, rotate=0]{$3\sigma$};
\definecolor{c}{rgb}{0.6,0.6,0};
\draw [c,dash pattern=on 0.80pt off 1.60pt ,line width=5.4] (14.1725,9.05244) -- (14.9775,9.05244);
\definecolor{c}{rgb}{0,0,0};
\draw [anchor=base west] (15.15,7.86005) node[scale=1.29711, color=c, rotate=0]{$5\sigma$};
\definecolor{c}{rgb}{0.8,0.2,0.2};
\draw [c,line width=5.4] (14.1725,8.07906) -- (14.9775,8.07906);
\definecolor{c}{rgb}{0,0,0};
\draw [anchor=base west] (14.05,6.74553) node[scale=1.42947, color=c, rotate=0]{$\mathcal{L}_{int}$ = 10 fb$^{-1}$};
\draw [anchor=base west] (14.05,5.57747) node[scale=1.42947, color=c, rotate=0]{$\gamma_{\mathcal{H}}$ = 0};
\end{tikzpicture}}\\	
	\scalebox{0.31}{\input{sigAllCan_Lumi100_LHPol.tex}}
	\scalebox{0.31}{\input{sigAllCan_Lumi100_UnPol.tex}}\\
	\caption{\label{fig:LamdaMass} The first row represents  attainable $\Lambda$ dependence on  $M_{u^*}$ for $\mathcal{L}_{int} = 10\; fb^{-1}$ and  $\sqrt{s} = 9.1$ TeV.  The second row shows the same relations for the same center of mass energy and  $\mathcal{L}_{int} = 100\; fb^{-1}$.}
\end{figure}

\begin{figure}[h!]
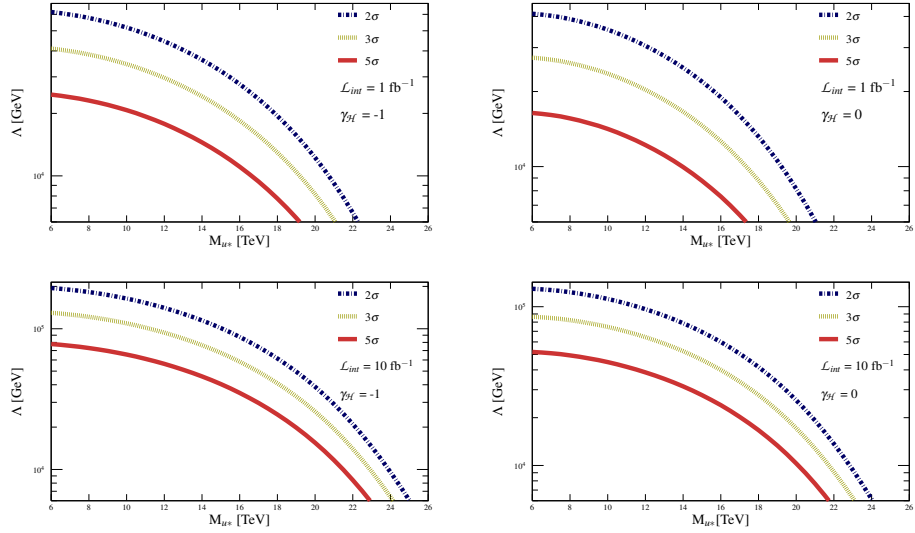

	\centering
	\scalebox{0.31}{\input{sigAllCan_Lumi1_CM29_LHPol.tex}}
		\scalebox{0.31}{\begin{tikzpicture}
\pgfdeclareplotmark{cross} {
\pgfpathmoveto{\pgfpoint{-0.3\pgfplotmarksize}{\pgfplotmarksize}}
\pgfpathlineto{\pgfpoint{+0.3\pgfplotmarksize}{\pgfplotmarksize}}
\pgfpathlineto{\pgfpoint{+0.3\pgfplotmarksize}{0.3\pgfplotmarksize}}
\pgfpathlineto{\pgfpoint{+1\pgfplotmarksize}{0.3\pgfplotmarksize}}
\pgfpathlineto{\pgfpoint{+1\pgfplotmarksize}{-0.3\pgfplotmarksize}}
\pgfpathlineto{\pgfpoint{+0.3\pgfplotmarksize}{-0.3\pgfplotmarksize}}
\pgfpathlineto{\pgfpoint{+0.3\pgfplotmarksize}{-1.\pgfplotmarksize}}
\pgfpathlineto{\pgfpoint{-0.3\pgfplotmarksize}{-1.\pgfplotmarksize}}
\pgfpathlineto{\pgfpoint{-0.3\pgfplotmarksize}{-0.3\pgfplotmarksize}}
\pgfpathlineto{\pgfpoint{-1.\pgfplotmarksize}{-0.3\pgfplotmarksize}}
\pgfpathlineto{\pgfpoint{-1.\pgfplotmarksize}{0.3\pgfplotmarksize}}
\pgfpathlineto{\pgfpoint{-0.3\pgfplotmarksize}{0.3\pgfplotmarksize}}
\pgfpathclose
\pgfusepathqstroke
}
\pgfdeclareplotmark{cross*} {
\pgfpathmoveto{\pgfpoint{-0.3\pgfplotmarksize}{\pgfplotmarksize}}
\pgfpathlineto{\pgfpoint{+0.3\pgfplotmarksize}{\pgfplotmarksize}}
\pgfpathlineto{\pgfpoint{+0.3\pgfplotmarksize}{0.3\pgfplotmarksize}}
\pgfpathlineto{\pgfpoint{+1\pgfplotmarksize}{0.3\pgfplotmarksize}}
\pgfpathlineto{\pgfpoint{+1\pgfplotmarksize}{-0.3\pgfplotmarksize}}
\pgfpathlineto{\pgfpoint{+0.3\pgfplotmarksize}{-0.3\pgfplotmarksize}}
\pgfpathlineto{\pgfpoint{+0.3\pgfplotmarksize}{-1.\pgfplotmarksize}}
\pgfpathlineto{\pgfpoint{-0.3\pgfplotmarksize}{-1.\pgfplotmarksize}}
\pgfpathlineto{\pgfpoint{-0.3\pgfplotmarksize}{-0.3\pgfplotmarksize}}
\pgfpathlineto{\pgfpoint{-1.\pgfplotmarksize}{-0.3\pgfplotmarksize}}
\pgfpathlineto{\pgfpoint{-1.\pgfplotmarksize}{0.3\pgfplotmarksize}}
\pgfpathlineto{\pgfpoint{-0.3\pgfplotmarksize}{0.3\pgfplotmarksize}}
\pgfpathclose
\pgfusepathqfillstroke
}
\pgfdeclareplotmark{newstar} {
\pgfpathmoveto{\pgfqpoint{0pt}{\pgfplotmarksize}}
\pgfpathlineto{\pgfqpointpolar{44}{0.5\pgfplotmarksize}}
\pgfpathlineto{\pgfqpointpolar{18}{\pgfplotmarksize}}
\pgfpathlineto{\pgfqpointpolar{-20}{0.5\pgfplotmarksize}}
\pgfpathlineto{\pgfqpointpolar{-54}{\pgfplotmarksize}}
\pgfpathlineto{\pgfqpointpolar{-90}{0.5\pgfplotmarksize}}
\pgfpathlineto{\pgfqpointpolar{234}{\pgfplotmarksize}}
\pgfpathlineto{\pgfqpointpolar{198}{0.5\pgfplotmarksize}}
\pgfpathlineto{\pgfqpointpolar{162}{\pgfplotmarksize}}
\pgfpathlineto{\pgfqpointpolar{134}{0.5\pgfplotmarksize}}
\pgfpathclose
\pgfusepathqstroke
}
\pgfdeclareplotmark{newstar*} {
\pgfpathmoveto{\pgfqpoint{0pt}{\pgfplotmarksize}}
\pgfpathlineto{\pgfqpointpolar{44}{0.5\pgfplotmarksize}}
\pgfpathlineto{\pgfqpointpolar{18}{\pgfplotmarksize}}
\pgfpathlineto{\pgfqpointpolar{-20}{0.5\pgfplotmarksize}}
\pgfpathlineto{\pgfqpointpolar{-54}{\pgfplotmarksize}}
\pgfpathlineto{\pgfqpointpolar{-90}{0.5\pgfplotmarksize}}
\pgfpathlineto{\pgfqpointpolar{234}{\pgfplotmarksize}}
\pgfpathlineto{\pgfqpointpolar{198}{0.5\pgfplotmarksize}}
\pgfpathlineto{\pgfqpointpolar{162}{\pgfplotmarksize}}
\pgfpathlineto{\pgfqpointpolar{134}{0.5\pgfplotmarksize}}
\pgfpathclose
\pgfusepathqfillstroke
}
\definecolor{c}{rgb}{1,1,1};
\draw [color=c, fill=c] (0,0) rectangle (20,11.6806);
\draw [color=c, fill=c] (2,1.16806) rectangle (18,10.5125);
\definecolor{c}{rgb}{0,0,0};
\draw [c,line width=0.9] (2,1.16806) -- (2,10.5125) -- (18,10.5125) -- (18,1.16806) -- (2,1.16806);
\definecolor{c}{rgb}{1,1,1};
\draw [color=c, fill=c] (2,1.16806) rectangle (18,10.5125);
\definecolor{c}{rgb}{0,0,0};
\draw [c,line width=0.9] (2,1.16806) -- (2,10.5125) -- (18,10.5125) -- (18,1.16806) -- (2,1.16806);
\draw [c,line width=0.9] (2,1.16806) -- (18,1.16806);
\draw (10,0.327056) node[scale=1.56475, color=c, rotate=0]{M$_{u*}$ [TeV]};
\draw [c,line width=0.9] (2,1.44839) -- (2,1.16806);
\draw [c,line width=0.9] (3.6,1.44839) -- (3.6,1.16806);
\draw [c,line width=0.9] (5.2,1.44839) -- (5.2,1.16806);
\draw [c,line width=0.9] (6.8,1.44839) -- (6.8,1.16806);
\draw [c,line width=0.9] (8.4,1.44839) -- (8.4,1.16806);
\draw [c,line width=0.9] (10,1.44839) -- (10,1.16806);
\draw [c,line width=0.9] (11.6,1.44839) -- (11.6,1.16806);
\draw [c,line width=0.9] (13.2,1.44839) -- (13.2,1.16806);
\draw [c,line width=0.9] (14.8,1.44839) -- (14.8,1.16806);
\draw [c,line width=0.9] (16.4,1.44839) -- (16.4,1.16806);
\draw [c,line width=0.9] (18,1.44839) -- (18,1.16806);
\draw [anchor=base] (2,0.782598) node[scale=0.900036, color=c, rotate=0]{6};
\draw [anchor=base] (3.6,0.782598) node[scale=0.900036, color=c, rotate=0]{8};
\draw [anchor=base] (5.2,0.782598) node[scale=0.900036, color=c, rotate=0]{10};
\draw [anchor=base] (6.8,0.782598) node[scale=0.900036, color=c, rotate=0]{12};
\draw [anchor=base] (8.4,0.782598) node[scale=0.900036, color=c, rotate=0]{14};
\draw [anchor=base] (10,0.782598) node[scale=0.900036, color=c, rotate=0]{16};
\draw [anchor=base] (11.6,0.782598) node[scale=0.900036, color=c, rotate=0]{18};
\draw [anchor=base] (13.2,0.782598) node[scale=0.900036, color=c, rotate=0]{20};
\draw [anchor=base] (14.8,0.782598) node[scale=0.900036, color=c, rotate=0]{22};
\draw [anchor=base] (16.4,0.782598) node[scale=0.900036, color=c, rotate=0]{24};
\draw [anchor=base] (18,0.782598) node[scale=0.900036, color=c, rotate=0]{26};
\draw [c,line width=0.9] (2,10.5125) -- (18,10.5125);
\draw [c,line width=0.9] (2,10.2322) -- (2,10.5125);
\draw [c,line width=0.9] (3.6,10.2322) -- (3.6,10.5125);
\draw [c,line width=0.9] (5.2,10.2322) -- (5.2,10.5125);
\draw [c,line width=0.9] (6.8,10.2322) -- (6.8,10.5125);
\draw [c,line width=0.9] (8.4,10.2322) -- (8.4,10.5125);
\draw [c,line width=0.9] (10,10.2322) -- (10,10.5125);
\draw [c,line width=0.9] (11.6,10.2322) -- (11.6,10.5125);
\draw [c,line width=0.9] (13.2,10.2322) -- (13.2,10.5125);
\draw [c,line width=0.9] (14.8,10.2322) -- (14.8,10.5125);
\draw [c,line width=0.9] (16.4,10.2322) -- (16.4,10.5125);
\draw [c,line width=0.9] (18,10.2322) -- (18,10.5125);
\draw [c,line width=0.9] (2,1.16806) -- (2,10.5125);
\draw (0.56,5.84029) node[scale=1.56475, color=c, rotate=90]{$\Lambda$ [GeV]};
\draw [c,line width=0.9] (2.24,1.16882) -- (2,1.16882);
\draw [c,line width=0.9] (2.24,1.88283) -- (2,1.88283);
\draw [c,line width=0.9] (2.24,2.50134) -- (2,2.50134);
\draw [c,line width=0.9] (2.24,3.0469) -- (2,3.0469);
\draw [c,line width=0.9] (2.48,3.53492) -- (2,3.53492);
\draw [anchor= east] (1.844,3.53492) node[scale=0.900036, color=c, rotate=0]{$10^{4}$};
\draw [c,line width=0.9] (2.24,6.74553) -- (2,6.74553);
\draw [c,line width=0.9] (2.24,8.62361) -- (2,8.62361);
\draw [c,line width=0.9] (2.24,9.95613) -- (2,9.95613);
\draw [c,line width=0.9] (18,1.16806) -- (18,10.5125);
\draw [c,line width=0.9] (17.76,1.16882) -- (18,1.16882);
\draw [c,line width=0.9] (17.76,1.88283) -- (18,1.88283);
\draw [c,line width=0.9] (17.76,2.50134) -- (18,2.50134);
\draw [c,line width=0.9] (17.76,3.0469) -- (18,3.0469);
\draw [c,line width=0.9] (17.52,3.53492) -- (18,3.53492);
\draw [c,line width=0.9] (17.76,6.74553) -- (18,6.74553);
\draw [c,line width=0.9] (17.76,8.62361) -- (18,8.62361);
\draw [c,line width=0.9] (17.76,9.95613) -- (18,9.95613);
\definecolor{c}{rgb}{0,0,0.4};
\draw [c,dash pattern=on 4.00pt off 2.40pt on 0.80pt off 2.40pt ,line width=5.4] (2,10.0722) -- (2.4,10.0345) -- (2.8,9.98551) -- (3.2,9.92521) -- (3.6,9.84073) -- (4,9.75003) -- (4.4,9.63868) -- (4.8,9.52598) -- (5.2,9.3906) -- (5.6,9.24124) --
 (6,9.07397) -- (6.4,8.89692) -- (6.8,8.70036) -- (7.2,8.49182) -- (7.6,8.26627) -- (8,8.02328) -- (8.4,7.76038) -- (8.8,7.47814) -- (9.2,7.17254) -- (9.6,6.84533) -- (10,6.50077) -- (10.4,6.12056) -- (10.8,5.71181) -- (11.2,5.27978) --
 (11.6,4.81992) -- (12,4.318) -- (12.4,3.78543) -- (12.8,3.2084) -- (13.2,2.58463) -- (13.6,1.92969) -- (14,1.22054) -- (14.0276,1.16806);
\definecolor{c}{rgb}{0.6,0.6,0};
\draw [c,dash pattern=on 0.80pt off 1.60pt ,line width=5.4] (2,8.19415) -- (2.4,8.15642) -- (2.8,8.10742) -- (3.2,8.04712) -- (3.6,7.96265) -- (4,7.87197) -- (4.4,7.76061) -- (4.8,7.6479) -- (5.2,7.51252) -- (5.6,7.36316) -- (6,7.1959) --
 (6.4,7.01884) -- (6.8,6.82227) -- (7.2,6.61373) -- (7.6,6.3882) -- (8,6.1452) -- (8.4,5.8823) -- (8.8,5.60008) -- (9.2,5.29448) -- (9.6,4.96724) -- (10,4.6227) -- (10.4,4.24248) -- (10.8,3.83375) -- (11.2,3.40169) -- (11.6,2.94182) -- (12,2.43992)
 -- (12.4,1.90735) -- (12.8,1.33032) -- (12.9041,1.16806);
\definecolor{c}{rgb}{0.8,0.2,0.2};
\draw [c,line width=5.4] (2,5.82805) -- (2.4,5.79032) -- (2.8,5.74133) -- (3.2,5.68102) -- (3.6,5.59655) -- (4,5.50586) -- (4.4,5.39451) -- (4.8,5.28179) -- (5.2,5.14641) -- (5.6,4.99706) -- (6,4.8298) -- (6.4,4.65271) -- (6.8,4.45618) --
 (7.2,4.24761) -- (7.6,4.02211) -- (8,3.77911) -- (8.4,3.51619) -- (8.8,3.23397) -- (9.2,2.92836) -- (9.6,2.60115) -- (10,2.2566) -- (10.4,1.87638) -- (10.8,1.46763) -- (11.0774,1.16806);
\definecolor{c}{rgb}{0,0,0};
\draw (10,11.301) node[scale=1.2177, color=c, rotate=0]{ };
\draw [anchor=base west] (15.15,9.80681) node[scale=1.29711, color=c, rotate=0]{$2\sigma$};
\definecolor{c}{rgb}{0,0,0.4};
\draw [c,dash pattern=on 4.00pt off 2.40pt on 0.80pt off 2.40pt ,line width=5.4] (14.1725,10.0258) -- (14.9775,10.0258);
\definecolor{c}{rgb}{0,0,0};
\draw [anchor=base west] (15.15,8.83343) node[scale=1.29711, color=c, rotate=0]{$3\sigma$};
\definecolor{c}{rgb}{0.6,0.6,0};
\draw [c,dash pattern=on 0.80pt off 1.60pt ,line width=5.4] (14.1725,9.05244) -- (14.9775,9.05244);
\definecolor{c}{rgb}{0,0,0};
\draw [anchor=base west] (15.15,7.86005) node[scale=1.29711, color=c, rotate=0]{$5\sigma$};
\definecolor{c}{rgb}{0.8,0.2,0.2};
\draw [c,line width=5.4] (14.1725,8.07906) -- (14.9775,8.07906);
\definecolor{c}{rgb}{0,0,0};
\draw [anchor=base west] (14.05,6.74553) node[scale=1.5883, color=c, rotate=0]{$\mathcal{L}_{int}$ = 1 fb$^{-1}$};
\draw [anchor=base west] (14.05,5.57747) node[scale=1.5883, color=c, rotate=0]{$\gamma_{\mathcal{H}}$ = 0};
\end{tikzpicture}}\\
	\scalebox{0.31}{\input{sigAllCan_Lumi10_CM29_LHPol.tex}}
	\scalebox{0.31}{\input{sigAllCan_Lumi10_UnPol.tex}}\\
	\caption{\label{fig:LamdaMass2} The first row represents  attainable $\Lambda$ dependence on  $M_{u^*}$ for $\mathcal{L}_{int} = 1\; fb^{-1}$ and  $\sqrt{s} = 28.8$ TeV.  The second row shows the same relations for the same center of mass energy and  $\mathcal{L}_{int} = 10\; fb^{-1}$.}
\end{figure}

We took compositeness scale equals $u^*$ mass or some specific values as 10, 15 and 30 TeV until this subsection. At this point, we scanned both the compositeness scale values and  $u^*$ mass for  discovery ($5\sigma$), observation ($3\sigma$) and exclusion ($2\sigma$) mass limits.  It can be clearly noticed from Figures \ref{fig:LamdaMass} and \ref{fig:LamdaMass2} that the higher compositeness scales correspond to the lower $u^*$ mass values. As it was expected, when the center of mass energy reached the 28.8 TeV with the highest luminosity value, the compositeness scale values had risen to the highest level for all $u^*$ mass spectra. Furthermore, the photon beam polarization will afford an opportunity to probe bigger compositeness scale values than the unpolarized photon beam-proton collision. 


In Tables \ref{tab:compLow} and \ref{tab:compHigh}, we summarize the highest attainable compositeness scale quantities for various $M_{u^*}$ values at the highest integrated luminosity values for both $\gamma p$ collider options. It is clearly seen that when  the photon beam polarization is in charge, compositeness scale values increase for the whole $M_{u^*}$ values. To illustrate, when we checked the compositeness scale values for $\sqrt{s} = 9.1$ TeV collider option with $M_{u^*} = 6$ TeV, the $\Lambda$ value increased to 70.5 TeV from 48.7 TeV at the $5\sigma$ significance.  Similarly, the compositeness scale value rose to 77.9 TeV from 51.9 TeV for $\sqrt{s} = 28.8 $ TeV collider option with the same $u^*$ mass values at the  $5\sigma$ significance.
\begin{table}[h!]
	\caption{Attainable top $\Lambda$ limits for  $M_{u^*}$ with the $\mathcal{L}_{int} = 100 \;\;fb^{-1}$.}
	\label{tab:compLow}
	
	\begin{tabular*}{\columnwidth}{@{\extracolsep{\fill}}cl|l|l|l|l@{}}
		\hline
		\multicolumn{2}{c|}{CM (TeV)} & \multicolumn{4}{c}{9.1} \\ \hline
		\multicolumn{2}{c|}{$\gamma_{\mathcal{H}}$} & \multicolumn{2}{c|}{-1} & \multicolumn{2}{c}{0} \\ \hline
		\multicolumn{2}{c|}{$M_{u^*}$ (TeV)} & \multicolumn{1}{c|}{6} & \multicolumn{1}{c|}{7} & \multicolumn{1}{c|}{6} & \multicolumn{1}{c}{7} \\ \hline
		\multirow{3}{*}{\begin{tabular}[c]{@{}c@{}}$\Lambda$ (TeV)\end{tabular}} & $5\sigma$& 70.5  & 30.8  & 48.7   & 21.2  \\ \cline{2-6} 
		& $3\sigma$ & 117  & 51.3  & 81.2 & 35.4  \\ \cline{2-6} 
		& $2\sigma$ & 176 & 76.9 & 122  & 53.1 \\  \hline
	\end{tabular*}
	
\end{table}

\begin{table}[h!]
	\caption{Attainable top $\Lambda$ limits for  $M_{u^*}$ with the $\mathcal{L}_{int} = 10 \;\;fb^{-1}$.}
	\label{tab:compHigh}
	
	\begin{tabular*}{\columnwidth}{@{\extracolsep{\fill}}cl|l|l|l|l|l|l@{}}
		\hline
		\multicolumn{2}{c|}{CM (TeV)} & \multicolumn{6}{c}{28.8} \\ \hline
		\multicolumn{2}{c|}{$\gamma_{\mathcal{H}}$} & \multicolumn{3}{c|}{-1} & \multicolumn{3}{c}{0} \\ \hline
		\multicolumn{2}{c|}{$M_{u^*}$ (TeV)} & 6 & 10 &15 &6 & 10 & 15 \\ \hline
		\multirow{3}{*}{\begin{tabular}[c]{@{}c@{}}$\Lambda$ (TeV)\end{tabular}} & $5\sigma$& 77.9  & 65.6  & 40.5 &51.9   &44.8 &27.7   \\ \cline{2-8} 
		& $3\sigma$ & 130  & 109 & 67.4  & 86.5   &74.7 & 46.2  \\ \cline{2-8} 
		& $2\sigma$ & 195 & 164 & 101 & 130 & 112&69.4\\  \hline
	\end{tabular*}
	
\end{table}

\subsubsection{Determination of the Chirality Structure of the $q^*$-$q$-$\gamma$ Vertex }

The FCC-pp collider option will afford an opportunity to investigate $M_{u^*}$  up to 50 TeV  mass limit \cite{akay2017} which essentially exceeds potential capacity of $\gamma p$ collider options. However, the $q^*$-$q$-$\gamma$ vertex could not be determined because the proton beams are  unpolarized. The FCC based $\gamma p$ colliders  have capability to handle  polarized photon beam which will allow to determine chirality structure of the excited quark interactions. Afterward, we executed asymmetry calculations taking compositeness scales equal  $u^*$ mass for $\eta_{L} =1$, $\eta_{R} = 0$  and $\eta_{L} =0$, $\eta_{R} = 1$  choices ($\eta_{L}$ and $\eta_{R}$ are  chirality factors in Equation \ref{eq:intLag}). Chirality structure of the $q^*$-$q$-$\gamma$ vertex are distinguished by looking at the asymmetry numbers given in  Table \ref{tab:asymmetry}.  Asymmetry calculation is done by Equation \ref{eq:asymmetry}:
\begin{equation}
\label{eq:asymmetry}
\mathcal{A} = \frac{\sigma(\gamma_{\mathcal{H}} = 1) - \sigma(\gamma_{\mathcal{H}} = -1) }{\sigma(\gamma_{\mathcal{H}} = 1) + \sigma(\gamma_{\mathcal{H}} = -1)}
\end{equation}
where $\mathcal{A}$ denotes asymmetry, $\sigma(\gamma_{\mathcal{H}} = -1)$ corresponds to the cross section numbers with helicity equal to -1 and $\sigma(\gamma_{\mathcal{H}} = 1)$ represents to cross section numbers with helicity equal to 1.    

\begin{table}[h!]
	\caption{\label{tab:asymmetry}  The polarization asymmetry for the excited $u$ quark} 
	\begin{tabular*}{\columnwidth}{@{\extracolsep{\fill}}llllllllllllll@{}}
		\hline 
		\begin{tabular}[c]{@{}c@{}}CM \\ (TeV)\end{tabular} & \begin{tabular}[c]{@{}c@{}}$M_{u^*}$ \\(TeV)\end{tabular} & $\gamma_{\mathcal{H}}$ & \multicolumn{2}{c|}{\begin{tabular}[c]{@{}c@{}} $\eta_{L} = 1, \eta_{R} = 0$ \end{tabular}}  & \multicolumn{2}{c}{\begin{tabular}[c]{@{}c@{}}$\eta_{L} = 0, \eta_{R} = 1$ \end{tabular}}\\ \hline
		& & &$\sigma$ (pb) & $\mathcal{A}$ &$\sigma$ (pb) & $\mathcal{A}$  \\ \hline
		\multirow{4}{*}{9.1} & \multirow{2}{*}{6} & -1 & 4.15$\times 10^{-2}$& \multirow{2}{*}{-0.99} & 8.07$\times 10^{-5}$  & \multirow{2}{*}{0.99} \\ \cline{3-4} \cline{6-6}
		&  & $\;\,$1 & 1.71$\times 10^{-4}$ &  &2.29$\times 10^{-2}$  &  \\  \cline{2-7} 
		& \multirow{2}{*}{7} & -1 & 6.50$\times 10^{-3}$  & \multirow{2}{*}{-0.98} & 2.78$\times 10^{-5}$ & \multirow{2}{*}{0.98} \\ \cline{3-4}\cline{6-6}
		&  & $\;\,$1 & 5.89$\times 10^{-5}$ &  &3.54$\times 10^{-3}$  &  \\ \hline
		\multirow{4}{*}{28.8} & \multirow{2}{*}{10} & -1 & 1.39$\times 10^{-1}$& \multirow{2}{*}{-0.99}  & 4.34$\times 10^{-4}$ & \multirow{2}{*}{0.99} \\ \cline{3-4} \cline{6-6}
		&  & $\;\,$1 & 9.20$\times 10^{-4}$ &  & 7.61$\times 10^{-2}$ &  \\ \cline{2-7} 
		& \multirow{2}{*}{15} & -1 & 2.23$\times 10^{-2}$ & \multirow{2}{*}{-0.99}& 3.56$\times 10^{-5}$  & \multirow{2}{*}{0.99} \\ \cline{3-4} \cline{6-6}
		&  & $\;\,$1 & 7.54$\times 10^{-5}$ &  &1.23$\times 10^{-2}$  &  \\ \hline 
	\end{tabular*}
\end{table}

\section{\label{sec:V}Conclusion}

In this work, we  analyzed  resonance production of the excited $u$ quark at the FCC based $\gamma p$ colliders that offer two possibilities: $\sqrt{s}^{max}_{\gamma p} = 9.1$ TeV with $\mathcal{L}_{int}$= 10-100 $fb^{-1}$ (ILC$\otimes$FCC ) and $\sqrt{s}^{max}_{\gamma p} = 28.8$ TeV with $\mathcal{L}_{int}$= 1-10 $fb^{-1}$ (PWFA-LC$\otimes$FCC ). It should be noted that at this stage, we did not consider hadronization and detector effects which may lead  to some decrease of discovery limits on $u^*$ mass and compositeness scale.

We conducted calculation of the $u^*$ mass limits for discovery ($5\sigma$), observation ($3\sigma$) and exclusion ($2\sigma$) confidence levels at the 10, 15, 30 TeV compositeness scales and at $\Lambda = M_{u^*}$, but the highest mass limits are achieved by taking $M_{u^*}$ equals  $\Lambda$. As seen from  Tables \ref{tab:lumMass10} and \ref{tab:lumMass32}, the photon beam polarization increases the mass limits for all confidence levels.  For $\gamma_{\mathcal{H}} = -1$, $\Lambda = M_{u^*}$ and $\mathcal{L}_{int} = 100 \;fb^{-1}$, attainable mass limits are  7.99 TeV for $5\sigma$, 8.23 TeV for $3\sigma$ and 8.40 TeV for $2\sigma$ at $\sqrt{s} = 9.1$ TeV collider option. Concerning the highest center of mass energy collider option ($\sqrt{s} = 28.8$ TeV),  the biggest attainable mass limits become  19.4 TeV for $5\sigma$, 21.1 TeV for $3\sigma$ and 22.2 TeV for $2\sigma$ confidence levels. To address these findings, ATLAS and CMS excluded $M_{u^*}$ up to 6 TeV with $37\;fb^{-1}$ integrated luminosity at $\sqrt{s} = 13$ TeV (with $\sqrt{s} = 14$ TeV and  $\mathcal{L}_{int} = 300 \;fb^{-1}$ this limit will potentially increase to $M_{u^*} = 7.5$ TeV). Therefore, the FCC-$\gamma p$ collider essentially superiors (3 times) the LHC potential.

Besides the specific values of the compositeness scale, we scanned the compositeness scale with respect to $M_{u^*}$. Our calculation results show that the highest compositeness scale  value is provided by the photon beam polarization (see Tables \ref{tab:compLow} and \ref{tab:compHigh}). Compositeness scale values are  evaluated as 77.9 TeV for $5\sigma$, 130 TeV for $3\sigma$ and 195 TeV for $2\sigma$ at the $\sqrt{s} = 28.8 $ TeV with  $\mathcal{L}_{int} = 10 \;fb^{-1}$, $M_{u^*} = 6 $ TeV and  $\gamma_{\mathcal{H}} = -1$. These values essentially exceed the LHC potential but the FCC-pp is much higher \cite{akay2017}.

Finally, if the excited quarks mass lies in the region mentioned above, the FCC-pp collider will apparently discover  $u^*$ before the construction of FCC based $\gamma p$ colliders. However, as seen from this study, latter ones will provide unique opportunity to determine chirality structure of $u^*$-$u$-$\gamma$ vertex by using the polarized photon beam.

\section{Acknowledgements}	
This study is supported by TUBITAK under the grant No: 114F337. We  thank Professor Yasar Onel for his support and contribution for encouraging such a research.

\bibliographystyle{unsrt}
\bibliography{excited_quark}
\end{document}